\documentclass[10pt,preprint]{aastex61}

\usepackage{hyperref}
\usepackage{color,amsmath}
\usepackage{soul}
\usepackage{graphicx}
\usepackage{times}
\usepackage[normalem]{ulem}

\newcommand{\comment}[1]{}
\renewcommand\sout{\bgroup \color{red} \ULdepth=-.5ex \ULset}
\def\simge{\mathrel{\rlap{\raise 0.511ex
     \hbox{$>$}}{\lower 0.511ex \hbox{$\sim$}}}}
\def\simle{\mathrel{\rlap{\raise 0.511ex
      \hbox{$<$}}{\lower 0.511ex \hbox{$\sim$}}}}
\newcommand{\kF}{k_\mathrm{{\scriptscriptstyle F}}}
\newcommand{\EUG}{E_\mathrm{UG}}
\newcommand{\EPNM}{E_\mathrm{PNM}}
\newcommand{\KPNM}{K_n}
\newcommand{\ESNM}{E_\mathrm{SNM}}
\newcommand{\Ksym}{K_\mathrm{sym}}
\newcommand{\EF}{E_\mathrm{F}}
\newcommand{\SLB}{S^\mathrm{LB}}

\newcommand{\mev}{\,\textrm{MeV}}

\begin{document}

\title{Symmetry Parameter Constraints From A Lower Bound On The Neutron-Matter Energy}

\author{Ingo Tews}
\email{itews@uw.edu}
\affiliation{
Institute for Nuclear Theory, University of Washington, Seattle, WA 98195-1550, USA}
\affiliation{JINA-CEE, Michigan State University, East Lansing, MI, 48823, USA}
\author{James M. Lattimer}
\email{james.lattimer@stonybrook.edu}
\affiliation{
Department of Physics and Astronomy, Stony Brook University, Stony Brook, NY 11794-3800, USA}
%%%%%
\author{Akira Ohnishi}
\email{ohnishi@yukawa.kyoto-u.ac.jp}
\affiliation{Yukawa Institute for Theoretical Physics, Kyoto University,
Kyoto 606-8502, Japan}
\author{Evgeni E. Kolomeitsev}
\email{e.kolomeitsev@gsi.de}
\affiliation{
Faculty of Natural Sciences, Matej Bel University,
Tajovskeho 40, SK-97401 Banska Bystrica, Slovakia}
\affiliation{Joint Institute for Nuclear Research, 
RU-141980 Dubna, Moscow Region, Russia}
%\homepage[]{Your web page}
%\thanks{}
%\altaffiliation{}
%%%%%

\received{2017 June 30}
\revised{2017 September 6}
\accepted{2017 September 17}
\published{2017 October 20}

\begin{abstract}

We propose the existence of a lower bound on the energy of pure neutron matter (PNM) on the basis of unitary-gas considerations. We discuss its justification from experimental studies of cold atoms as well as from theoretical studies of neutron matter. We demonstrate that this bound results in limits to the density-dependent symmetry energy, which is the difference between the energies of symmetric nuclear matter and PNM. In particular, this bound leads to a lower limit to the volume symmetry energy parameter $S_0$. In addition, for assumed values of $S_0$ above this minimum, this bound implies both upper and lower limits to the symmetry energy slope parameter $L$, which describes the lowest-order density dependence of the symmetry energy. A lower bound on the neutron-matter incompressibility is also obtained. These bounds are found to be consistent with both recent calculations of the energies of PNM and constraints from nuclear experiments. Our results are significant because several equations of state that are currently used in astrophysical simulations of supernovae and neutron star mergers, as well as in nuclear physics simulations of heavy-ion collisions, have symmetry energy parameters that violate these bounds. Furthermore, below the nuclear saturation density, the bound on neutron-matter energies leads to a lower limit to the density-dependent symmetry energy, which leads to upper limits to the nuclear surface symmetry parameter and the neutron-star crust-core boundary. We also obtain a lower limit to the neutron-skin thicknesses of neutron-rich nuclei. Above the nuclear saturation density, the bound on neutron-matter energies also leads to an upper limit to the symmetry energy, with implications for neutron-star cooling via the direct Urca process.
\end{abstract}

%\maketitle
\section{Introduction}
The nuclear symmetry energy $(S(u))$ is one of the decisive ingredients
in compact-star astrophysics as well as in nuclear physics. It provides 
the pressure of neutron-star matter, which is nearly pure neutron matter
(PNM) near the saturation density $n_0\simeq0.16$ fm$^{-3}$,  and
largely determines neutron-star radii~\citep{Lattimer01} and therefore
properties of their crusts, moments of inertia, tidal polarizabilities, and 
binding energies~\citep{Lattimer07}.  The symmetry energy is also
important in calculations of the r-process~\citep{Mumpower16}, 
supernovae~\citep{Fischer14}, and neutron-star 
mergers~\citep{Bauswein16}. Terrestrial experiments measuring nuclear
masses, dipole resonances, and neutron-skin thicknesses can constrain
the symmetry energy~\citep{Lattimer13}, as can experiments using 
normal and radioactive nuclear beams~\citep{Oertel17}.

The symmetry energy $S(u)$ can be obtained from the energy per
particle in nuclear matter, $E(u,x)$, where $x$ is the proton fraction
and $u=n/n_0$ is the density in units of saturation density $n_0$. The
energy per particle at a given density varies between a minimum,
symmetric nuclear matter (SNM, $x=0.5$), and a maximum, PNM ($x=0$). 
The symmetry energy is defined as the difference
of these energies: 
\begin{equation} 
S(u)=E(u,0)-E(u,1/2)\,.
\label{eq:sym}
\end{equation}
One can expand $E(u,x)$ in terms of the neutron 
excess $(1-2x)$. From the symmetry properties of nuclear matter, there 
are no terms with odd powers of the neutron excess, so the lowest-order 
term in the expansion is quadratic, i.e., 
\begin{equation}
E(u,x)= E(u,1/2)+S_2(u)(1-2x)^2+S_4(u)(1-2x)^4+\dots \,.
\label{eq:quad}\end{equation}
There is little experimental evidence concerning the magnitude of
quartic and higher-order terms because laboratory
nuclei are nearly symmetric.  Studies suggest that these
terms are small~\citep{Carbone2014,Wellenhofer2016}.
If the energy is divided into kinetic and potential
contributions, then the kinetic energy of degenerate non-interacting
fermions can be expanded as
\begin{align}
E_{\rm kin}(u,x)&={3\over5}{\hbar^2\over2m}\left(3\pi^2n_0u\right)^{2/3}\left[(1-x)^{5/3}+x^{5/3}\right]\\
&=T\left[1+{5\over9}(1-2x)^2+{5\over243}(1-2x)^4+\cdots\right],
\label{eq:kin}\end{align}
where $T=3\hbar^2(3\pi^2n_0)^{2/3}/(2^{2/3} 10m)\simeq22.1$ MeV.  Keeping 
instead only the quadratic term in Equation~\eqref{eq:quad}, one would obtain
\begin{equation}
E_{\rm kin,quad}=T\left[1+\left(2^{2/3}-1\right)(1-2x)^2\right],
\end{equation}
so the kinetic contributions to the symmetry energy to quadratic order would 
differ by
\begin{equation}
\Delta S_{\rm kin}=S_{\rm kin,quad}-S_{\rm kin}=T\left[2^{2/3}-{14\over9}\right]\simeq0.032T\,.
\end{equation}
This difference would have to be made up from quartic and higher-order 
terms. Indeed, the quartic contribution to the kinetic symmetry energy 
contributes $5T/243=0.021T$, about 2/3 of the missing amount.  

Most theoretical calculations of the potential contribution to the symmetry
energy also find only small corrections beyond the quadratic 
term~\citep{Carbone2014,Wellenhofer2016}. Recent 
calculations of neutron-rich matter~\citep{Drischler16} have shown that up 
to densities approaching $ u\sim1.5$ the quadratic assumption is accurate to 
better than 1~MeV in the symmetry energy per particle for all values of $x$.   
However, note that the kinetic contributions to quartic and 
higher-order terms vary as $u^{2/3}$ and potential contributions vary with 
higher powers of $u$. Thus when $u>>1$, the neglect of these terms may 
be unjustified. 

Keeping only the quadratic term for the entire range of proton
fractions, i.e., assuming $S(u)=S_2(u)$, the energy and pressure of
PNM can be expressed through the traditional symmetry energy
parameters
\begin{align}
S_0&={1\over8}\left({\partial^2E\over\partial x^2}\right)_{u=1}=(S)_{u=1},\\
L&={3\over8}\left(u{\partial^3E\over\partial u\partial x^2}\right)_{u=1}=3\left(u{dS\over du}\right)_{u=1},\\
K_{\rm sym}&={9\over8}\left(u^2{\partial^4E\over\partial u^2\partial x^2}\right)_{u=1}=9\left(u^2{d^2S\over du^2}\right)_{u=1},
\end{align}
and
\begin{align}
Q_{\rm sym}&={27\over8}\left(u^3{\partial^5E\over\partial u^3\partial x^2}\right)_{u=1}=27\left(u^3{d^3S\over du^3}\right)_{u=1}.
\label{eq:sympar}\end{align}
The symmetry energy can then be Taylor expanded around the
saturation density, $u=1$,
\begin{align}
%\hspace*{-.5cm}
S(u)=S_0+\frac{L}{3}(u-1)+\frac{\Ksym}{18}(u-1)^2+\frac{Q_{\rm sym}}{162}(u-1)^3
%=S_0\!+\!\frac{L}{3}\,(u\!-\!1)\!+\!\frac{\Ksym}{18}(u\!-\!1)^2%
+\mathcal{O}\left[(u-1)^4\right]\,.
\label{Eq:Sexp}
\end{align}
The most important parameters are $S_0$ and $L$, the symmetry
energy and its slope at $u=1$.  $K_{\rm sym}$ and $Q_{\rm sym}$ are the incompressibility and skewness parameters, respectively.

An abundance of
information concerning the allowed ranges for $S_0$ and $L$ exists from
nuclear experiments and theoretical studies of neutron matter.  In
particular, a strong positive correlation between their values can be
deduced in a nearly model-independent manner from nuclear binding
energies~\citep{Lattimer13}. $\Ksym$ describes the 
curvature of $S$ near $u=1$, and can be probed in a limited way using 
the giant monopole resonance, see \cite{Blaizot1980, Farine97} and \cite{Stone2015}.
$Q_{\rm sym}$ describes the skewness of $S$ near $u=1$, but it is extremely 
difficult to infer its value from experimental data. Thus, experimental and 
theoretical knowledge of $\Ksym$ and $Q_{\rm sym}$ is poor.

With knowledge of the binding energy of SNM at
saturation, $E_0\simeq-16\pm0.5$ MeV, $S_0$ determines the energy per
particle of PNM at the same density: $E(1,0)=S_0+E_0$.  In addition,
the pressure $p(u,x)$ of PNM is directly related to $L$: $p(1,0)=Ln_0/3$.  
PNM is a system of relevance for astrophysics because neutron-star matter
is very close to PNM due to the small proton fractions in $\beta-$
equilibrium near the nuclear saturation density. There have been many
microscopical determinations of the neutron-matter equation of state,
using a multitude of nuclear Hamiltonians and many-body methods.
These include, e.g., self-consistent Green's function (SCGF)
methods (see~\cite{SCGF2016} for recent results), Quantum Monte 
Carlo (QMC) calculations~\citep{Gezerlis10, Gandolfi12, Lynn:2015jua},
many-body perturbation theory (MBPT)
calculations~\citep{Hebeler10, Tews13}, and variational
methods~\citep{FP, APR, Togashi}.  Based on these calculations
one can obtain constraints on the symmetry energy parameters if
one assumes the quadratic expansion to be approximately valid.  The
extracted symmetry energy parameters are consistent with experimental
determinations~\citep{Lattimer13, Tews13}. Due to the intimate connection
of PNM with the symmetry energy, any additional information about PNM
translates into additional constraints on $S_0$ and $L$.  Establishing
a lower limit for the energy of PNM, e.g., would result in
general limits to the symmetry energy parameters that so far have not
been considered.

One possiblilty of a general lower limit could come from the unitary 
gas (UG). Universal behavior emerges for fermions interacting via
pairwise $s$-wave interactions with an infinite scattering length ($a_s$)
and a vanishing effective range ($r_\mathrm{eff}$), {\em i.e.} the UG (see Ref.~\citep{Zwierlein15} for an historical review). Since in this
case the average particle distance is the only length scale of the system,
the ground state energy per particle in the UG $\EUG$ is
proportional to the Fermi energy $\EF$, $\EUG = \frac35 \,\EF\,\xi_0$, 
where the Bertsch parameter $\xi_0$ has the
experimentally measured value of $\xi_0\simeq 
0.37$~\citep{Ku12,Zurn13}. 

PNM at very low densities, e.g. $n\sim 0.01 n_0$, where only $s$-wave 
contributions are important, is considered to show crossover 
behavior~\citep{Matsuo06} and to be close to the unitary limit since 
the $s$-wave scattering length of the $nn$ system is $a_s = -18.9$
fm~\citep{Machleidt2001}. This corresponds to $(a_s k_F )^{-1} 
\simeq -0.15 (-0.025)$ at $0.01n_0 (2 n_0)$, where $k_F = (3\pi^2 n)^{1/3}$ 
is the neutron Fermi momentum, whereas the UG limit is $(a_s k_F )^{-1} 
= 0$. However, at large densities, differences in the underlying interactions 
become important and lead to different effects in both systems. This includes
effective-range effects, interactions in higher partial waves, tensor contributions, 
etc. As an example, while the UG has a vanishing effective range, 
neutrons have an effective range of $r_\mathrm{eff}\sim 2.7$ fm~\citep{Machleidt2001}.

While it has been suggested before that the UG is lower in energy
than PNM, e.g.,~\cite{PTPSuppl2002, PTEP2012} and \cite
{ARNPS2015}, it has so far not been applied to obtain limits on the
symmetry energy parameters. We therefore follow the implications 
of making that conjecture,
\begin{equation}
E_{\rm {PNM}}(n)\ge \frac35 \xi_0 E_{\rm F}(n),\label{Eq:conjecture}
\end{equation}
at densities of $n\lesssim 1.5 n_0$.  We demonstrate that the nuclear 
symmetry energy parameters are thereby significantly constrained 
by the UG energy.

\section{The UG as a Lower Energy Limit to Pure Neutron Matter}

Nucleon-nucleon (NN) scattering can be described in terms of phase 
shifts in different partial waves. A positive phase shift in a particular
scattering channel indicates attractive interactions, while a negative
phase shift signals repulsive interactions in that channel. For neutrons 
at low energies, \cite{Schwinger47} showed that 
the phase shift $\delta$ can be related to the the neutron momentum
$k$ via the effective-range expansion, which is given for $s$-wave
interactions by
\begin{equation}
k \cot(\delta_S)=-\frac{1}{a_s}+\frac12 r_\mathrm{eff} k^2  + \mathcal{O}(k^4) \,,
\end{equation}
with the $s$-wave scattering length $a_s$ and the effective range 
$r_\mathrm{eff}$. 

Mathematically, the scattering length $a_s$ describes the slope of 
the phase shift at $k=0$. In particular, we use here the common 
convention that negative signs of $a_s$ correspond to positive slopes
and vice versa. Then a positive $s$-wave scattering length signals
the existence of a bound state, c.f., the nuclear $^3S_1$ partial 
wave, and a negative $a_s$ indicates no bound state, c.f., the 
$^1S_0$ partial wave. Starting from a negative scattering length,
increasing the attraction of the interaction leads to a growing $|a_s|$. 
If $|a_s|\to\infty$, then a bound state appears at the threshold. 

For neutron matter, $r_\mathrm{eff} \approx 2.7\, \rm{fm}$ and
$a_s=-18.9\, \rm{fm}$. Thus, the neutron-neutron interaction leads to
no bound state in the $s$-wave channel but is nevertheless strongly
attractive. At very low densities, $k_F \ll 1$, the interparticle spacing is 
much larger than the effective range of the interaction. The system is 
then fully described by $k_F$ and $a_s$. Systems with a similar 
scattering length $a_s$ and the same Fermi momentum $k_F$ will 
experience the same physics. If in addition $a_s$ can
be considered very large, or $(|a_s k_F|)^{-1}\to 0$, then the system is
completely described by the density or $k_F$. 
This regime is called the universal regime, and the system's energy 
must be proportional to the energy of a free Fermi gas, hence
\begin{equation}
E(n)= \xi_0 E_{\text{FG}}(n)\,.
\end{equation}
The UG, with $a_s\to \pm \infty$ and $r_\mathrm{eff}=0$, has 
universal behavior, but also will any dilute fermionic
system with $r_\mathrm{eff} \ll k_F^{-1} \ll |a_s|$, e.g., neutron matter at low
densities. In the unitary limit, the properties of the potential, e.g., its
shape, become irrelevant and the system is completely described by 
the dimensionless combination $a_s k_F$. 
$\xi_0$ can be measured in experiments with ultracold 
atoms~\citep{Ku12, Zurn13} around a Feshbach resonance.
In this resonance, $a_s$ can be tuned over several orders of 
magnitude by varying the magnetic field.

For any fermionic system, unitary or not, we define $\xi=E/E_\mathrm{FG}$.
\cite{Lee57} showed that in the limit of zero density, $k_F\to0$ 
and $(k_Fa_0)^{-1}\ll-1$, which is far from the unitary limit:
\begin{equation}
\xi=1+{10\over9\pi}k_Fa_0 +{(11-2\ln2)\over21\pi^2}(k_Fa_0)^2+\cdots.
\label{eq:LY}\end{equation}
Here, $a_0$ is the scattering length in the relevant channel.
Near unitarity, $|k_Fa_0|\to\infty$, cold atom experiments~\citep{Navon10} 
show that $\xi$ changes linearly with $(k_Fa_0)^{-1}$ for $|k_Fa_0|\simge50$; 
approximately
\begin{equation}
\xi\simeq\xi_0-0.93(k_Fa_0)^{-1}+\cdots,
\label{eq:xi0exp}\end{equation}
which also compares favorably with
theoretical predictions from QMC calculations of cold
atoms~\citep{PTEP2012}.
Thus, forces with a finite negative $a_s$ at very 
low densities have $\xi>\xi_0$ and higher energies than the UG.
Then for spin-one-half fermions interacting solely via the $s$-wave 
interactions that do not produce two-body bound states ($a_s<0$), the 
unitary limit, $a_s \to -\infty$ and $r_\mathrm{eff}\to 0$, indicates the
largest attraction.  At very low densities, when neutron matter can be
described solely by its $a_s$, the UG energy 
serves as a lower bound for $\EPNM$ because the magnitude of the
neutron $a_s$ is smaller, $|a_s|<\infty$, which indicates less
attraction~\citep{PTEP2012, ARNPS2015}.

As stated above, at higher densities $s$-wave effective-range effects
are not negligible anymore (e.g. $k_Fr_\mathrm{eff}=4.53$ at $n_0$).
However, a finite effective range reduces the attraction and increases
the energy per particle. For example, when $k_Fr_\mathrm{eff} \lesssim
0.35$, the energy increases with density as
$E=E_\mathrm{FG}\xi=E_\mathrm{FG}(\xi_0 + 0.12(3) k_F
r_\mathrm{eff}+\dots)$~\citep{ARNPS2015}. At higher densities, up to
at least $k_Fr_\mathrm{eff}=5$, correspondingly larger increases in
$\xi$ have been found~\citep{SchwenkPethick}. Because a finite
$(a_sk_F)^{-1}$ and a finite $r_\mathrm{eff}$ both lead to an increase
of the the effective Bertsch parameter $\xi$, the energy of
neutron matter is higher than for the UG, at least when only
realistic $s$-wave interactions are considered.

At even higher densities, $p$- and higher partial wave interactions 
may spoil this bound: the average $p$-wave interaction is very small but 
attractive, as are $d$-wave contributions to neutron matter.

When comparing the UG with neutron-matter calculations using
only the full NN interactions, one finds that the
energies of neutron matter for $n\simle n_0$ and the UG are
comparable although the underlying interactions are very
different. Softer (harder) NN interactions lead to slightly more (less)
attraction compared to the UG. This may be due
to fewer short-range correlations for softer interactions. In the
NN-only results of~\cite{Tews13}, using soft chiral Hamiltonians, the 
neutron-matter energy at $n_0$ was found to be approximately $2$ MeV 
below that of the UG.

Because we found that some soft Hamiltonians violate the UG bound on 
the NN-only level, we performed additional MBPT calculations at third order 
similar to \cite{Tews13}, using two soft Hamiltonians and including 
only $s$-wave, $s+p$-wave, and $s+p+d$-wave interactions. When 
considering only $s$-wave interactions we found that the energy per neutron 
lies $5-6$ MeV above the UG bound at $n_0$. This supports our justification 
for realistic $s$-wave interactions. For a finite $\ell$, we found that $p$- and 
$d$-wave interactions contribute attractively and lower the energy to $2$ 
MeV below the UG at $n_0$. Higher partial waves ($\ell>2$) add 
only a small contribution. 

\begin{figure}[t]
\centerline{
\includegraphics[width=0.6\textwidth,trim=0cm 0cm 0cm 0cm, clip=]{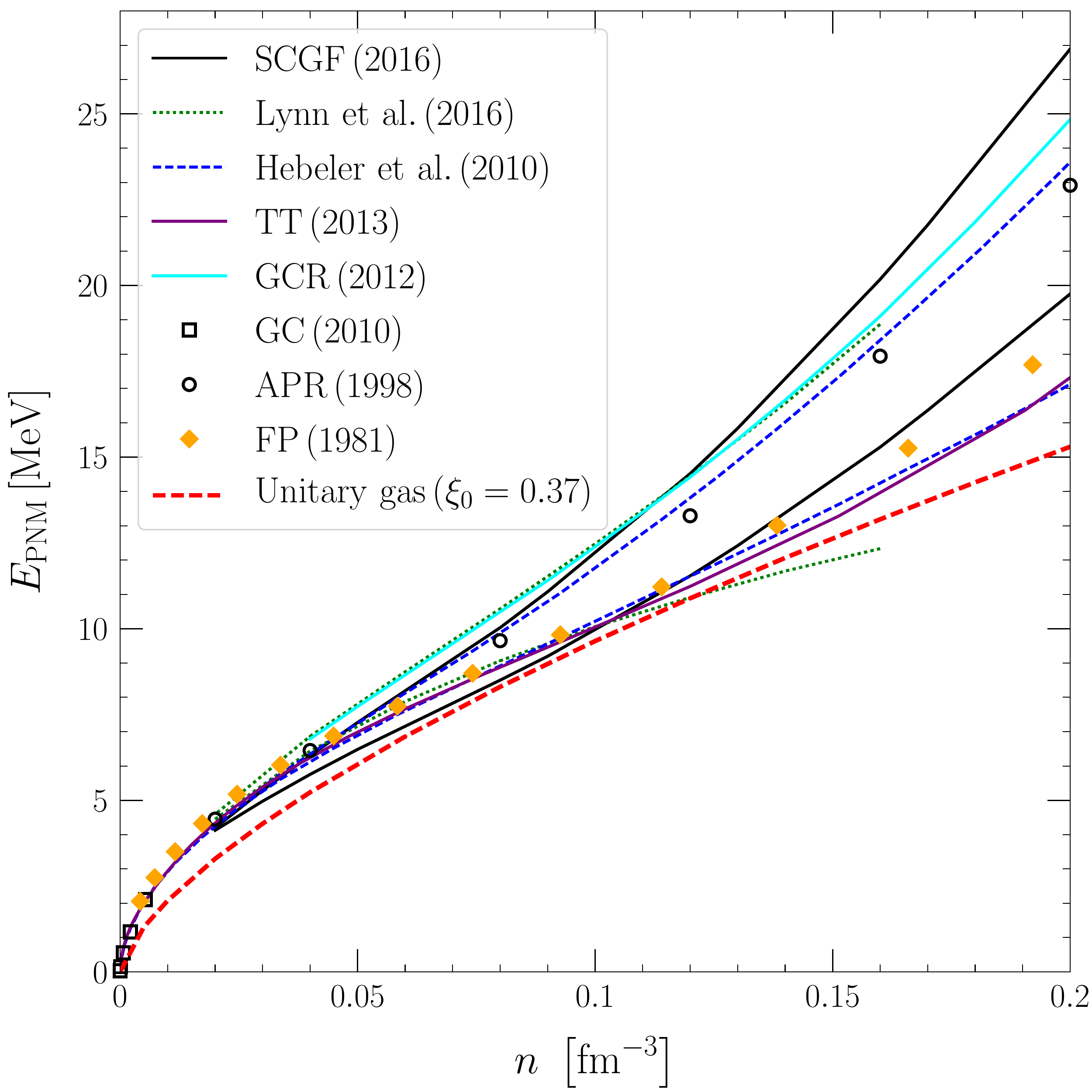}
}
\caption{UG bound with $\xi_0=0.37$ compared to ab initio calculations
of~\cite{SCGF2016} (SCGF), \cite{Lynn:2015jua} (Lynn et al.), 
\cite{Hebeler10} (Hebeler  et al.), \cite{Togashi} (TT), \cite{Gandolfi12} (GCR), \cite{Gezerlis10} (GC), \cite{APR} (APR), and \cite{FP} (FP).
}
\label{Fig:EFGbound}
\end{figure}

In neutron matter, however, several ab initio calculations have shown
the importance of three-body (3N) forces, which lead to, in net, additional 
repulsion, and increase the neutron-matter energy by several MeV at 
$n_0$. The inclusion of 3N forces is sufficient to more than compensate for the
$p$- and $d$-wave attraction, so that the neutron-matter energy
has values above the UG. In Figure~\ref{Fig:EFGbound} we show a
comparison of the UG bound with several ab initio
calculations for PNM, including both NN and 3N forces, of
\cite{SCGF2016,Lynn:2015jua, Togashi, Gandolfi12,Hebeler10,
  Gezerlis10,APR} and \cite{FP}. Only one calculation violates the
UG constraint within its uncertainty band: a QMC N$^2$LO calculation 
using soft chiral
forces~\citep{Lynn:2015jua}.  This is due to artifacts from
local regulators, which lead to less repulsion from 3N
forces~\citep{Tews:2015ufa, Dyhdalo:2016ygz}, and is peculiar to that 
interaction.

We emphasize the fact that ab initio calculations with realistic NN 
and 3N forces support our conjecture that $E_{\rm PNM}> E_{\rm UG}$.
Although the underlying interactions in both systems are rather different
and thus it cannot be strictly proven that the UG is a lower 
bound for the neutron-matter energy at all densities, the combination of 
effective-range effects, small $p$-wave and tensor interactions, and 
repulsive 3N forces in neutron matter strongly suggests that this
conjecture is justified. It is the consequence of a realistic nuclear 
Hamiltonian and is not altered by the differences in interactions and 
wave functions within the two systems.

\section{The Minimal Constraint on the Symmetry Energy\label{Sec:minimal}}

We now show what the inequality in Equation~\eqref{Eq:conjecture} implies for the 
symmetry energy parameters. At a baryon density of $n=u \, n_0$, the 
UG energy is
\begin{align}
\EUG(u) =& \frac{3}{5}\,\EF(u)\,\xi_0 = \frac{3\hbar^2 \kF^2}{10 m_n}\,\xi_0
= \EUG^{0}\,u^{2/3}
\ ,\label{Eq:EUG}
\end{align}
where $m_n$ is the neutron mass and $\EUG^{0}=\EUG(u\!=\!1)\simeq12.6$~MeV. The SNM energy ($\ESNM$) can be expanded as 
\begin{align}
\ESNM(u) =& E_0 + {K_0\over18}\,(u-1)^2 + {Q_0\over162}(u-1)^3+\mathcal{O}[(u-1)^4]
\ ,\label{Eq:ESNM}
\end{align}
where $E_0\simeq-16$ MeV and $K_0\simeq230$ MeV are the saturation 
energy and incompressibility parameters, respectively. The parameter 
$Q_0\sim-300$ MeV is the skewness parameter, whose value is not well 
known.  Using Equation~(\ref{eq:sym}) for the definition of the symmetry energy, 
the conjecture \eqref{Eq:conjecture} yields the lower bound $\SLB(u)$:
\begin{align}
S(u)
%\equiv&\EPNM(u)-\ESNM(u)
%\geq& \EUG(u)-\ESNM(u)
%\equiv S_\mathrm{LB}(n)
%\nonumber\\
\geq & \EUG^0u^{2/3}-\left[E_0+\frac{K_0}{18}\,(u-1)^2 +\frac{Q_0}{162}\,(u-1)^3\right]
\equiv \SLB(u)
\ .\label{Eq:S_n}
\end{align}

\begin{figure}[t]
\center\includegraphics[width=0.6\textwidth]{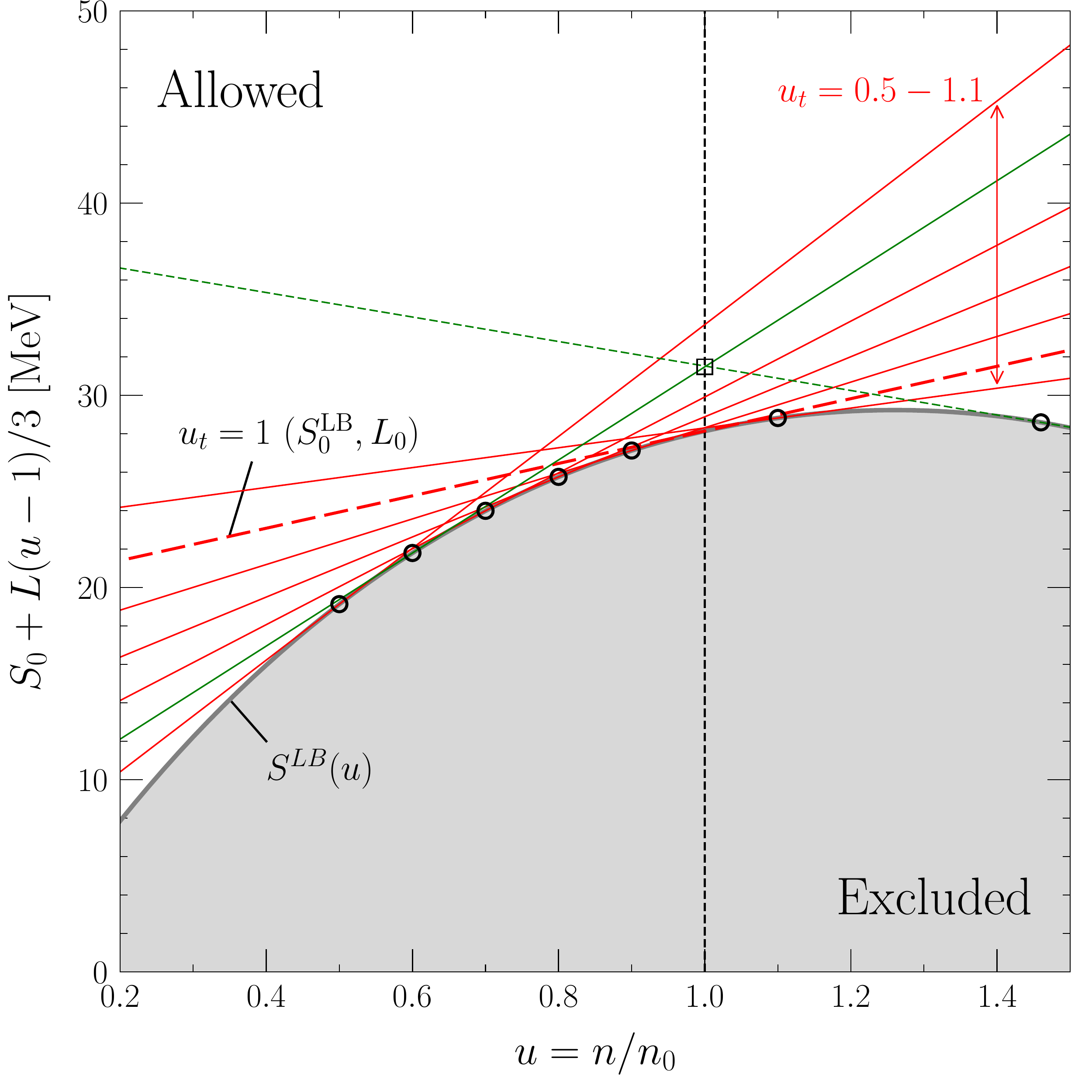}
\caption{UG bound on the symmetry energy,
using Equation~\eqref{Eq:S_n} and typical parameters as described in the text.
The shaded area  is the region excluded for $S_0 + L(u-1)/3.$.  The dashed red curve 
shows the tangent curve to the boundary $S^\mathrm{LB}_0$ at $u_t=1$, for 
which $S_0=S_0^\mathrm{LB}$; other solid lines
show the tangent curves for various values of $u_t$ (and $S_0$).  The green 
curves show one particular case for $S_0>S_0^\mathrm{LB}$: the solid 
green curve has $u_t= 0.6$ and the dotted green curve has $u_t>1$.
}
\label{Fig:S_n}
\end{figure}

In Figure~\ref{Fig:S_n}, we show the lower bound on the symmetry energy 
imposed by the 
UG constraint using typical values for the nuclear parameters 
($E_0=-16$ MeV, $K_0=230$ MeV, $Q_0=-300$ MeV, and 
$\EUG^0=12.6$ MeV).The shaded area shows the excluded region, inside which 
the symmetry energy should not enter. The boundary of this region is quite 
insensitive to the values of $K_0$ 
and $Q_0$: variations of $\Delta K_0=\pm30$ MeV and $\Delta Q_0=\pm300$ 
MeV each move the boundary at $u=0.2$ by only about $\pm1$ MeV, and less 
for values of $u$ closer to 1.

%%%%%%%%%%%%%%%%%%%%%%%%%%%%%%

%%%%%%%%%%%%%%%%%%%%%%%%%%%%%%%%%%%%%%%%%%%%%%%%%%%%%

From Equation~\eqref{Eq:S_n}, it is clear that $S_0 \equiv S(u\!=\!1)$ is bounded 
from below:
$S_0 \geq \EUG^0-E_0\equiv S_0^\mathrm{LB}$.
Because the symmetry energy should not enter the excluded area, at 
$S_0 = S_0^\mathrm{LB}$ the slopes of $S(u)$ and $\SLB(u)$ must 
agree, as shown by the tangential red-dashed line in Figure~\ref{Fig:S_n}.
We thus find
\begin{align}
L=3\left.u{d\SLB\over du}\right|_{u=1}=2\EUG^{0}\equiv L_0\ .
\label{Eq:l}
\end{align}
It also follows that the curvature of $S$ at $u=1$ must be greater than that of
$S^\mathrm{LB}$ (which is negative), or else $S$ could penetrate into the 
excluded region for $u\ne1$.  This implies
\begin{equation}
K_{\rm sym}\ge9\left.u^2{ d^2S^\mathrm{LB}\over du^2}\right|_{u=1}=-2\EUG^0-K_0\equiv K_{\rm sym,0}, \qquad\KPNM\ge-2\EUG^0\equiv K_{n,0},
\label{Eq:k}\end{equation}
where $\KPNM=K_0+K_{\rm sym}$ is the incompressibility of PNM at 
saturation density.  We also define the skewness 
of PNM at $u=1$, $Q_n=Q_0+Q_{\rm sym}$.
We find that $S_0^\mathrm{LB}, L_0$, and $K_{n,0}$ are 
independent of $K_0$ and $Q_0$, the most uncertain of the saturation 
parameters, and thus appear to be very general. 

By choosing the UG as a lower bound for neutron matter, we also pick a 
specific density dependence. We stress that a more realistic density
dependence for the lower bound would be steeper than $u^{2/3}$ and lead 
to a less concave symmetry energy\footnote{Equation~(\ref{eq:xi0exp}) shows that 
$d\xi/du>0$.}.  In turn, this would lead to even more stringent constraints. 
Thus by choosing the particular density dependence $u^{2/3}$, we are 
suggesting a more conservative bound and are underestimating our constraint.

Using the expansion for $S(u)$ of Equation~(\ref{Eq:Sexp}), allows
Equation~(\ref{Eq:S_n}) to be expressed as
\begin{align}
S_0+\frac{L}{3}(u-1)\geq
\EUG^0u^{2/3}-\left[E_0 + \frac{\KPNM}{18}(u-1)^2 +\frac{Q_n}{162}(u-1)^3\right]
\ .\label{Eq:ineq}
\end{align} 

%%%%%%%%%%%%%%%%%%%%%%%%%%%
\begin{figure}[h]
\begin{center}
\includegraphics[width=0.6\textwidth,trim= 0 0 0 0, clip=]{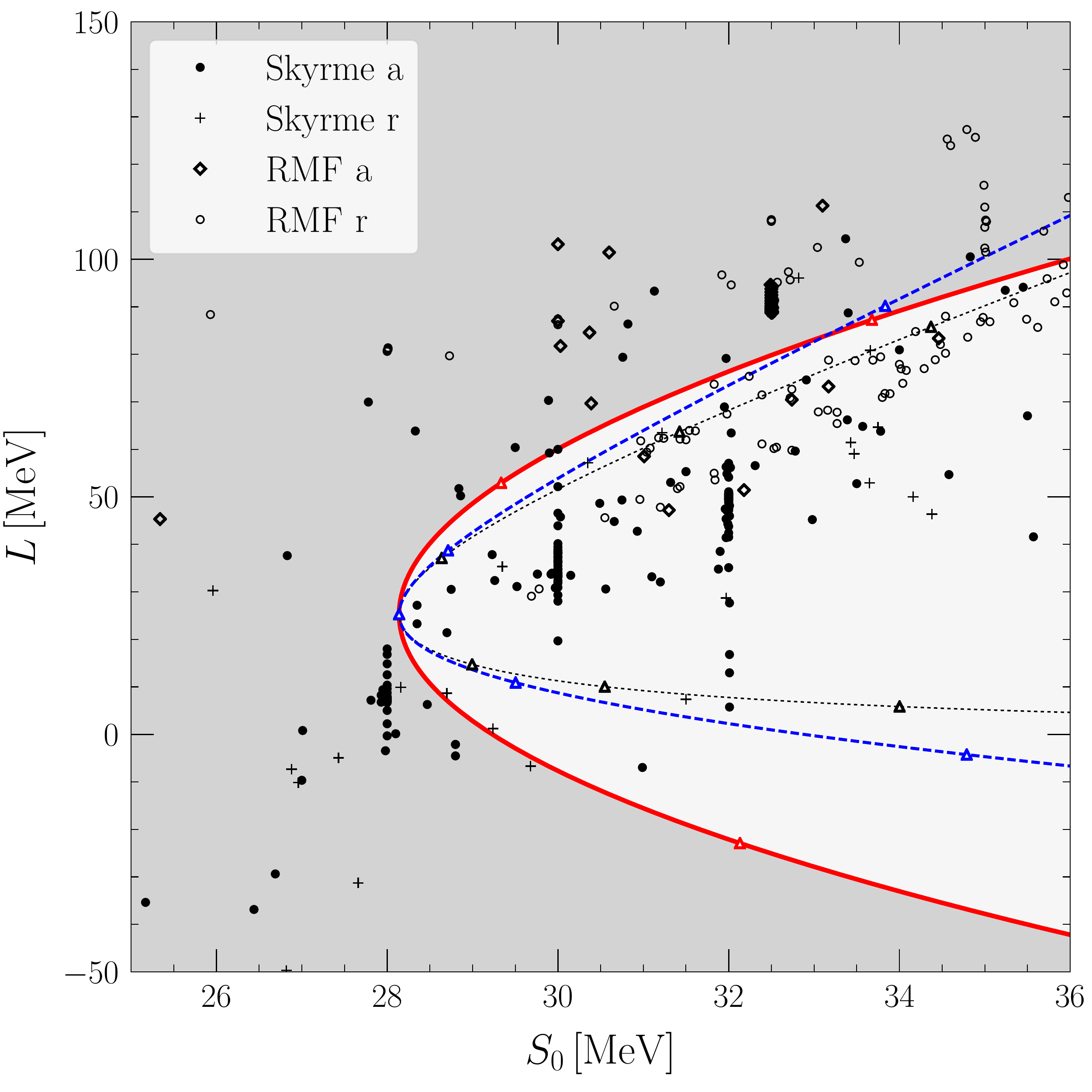}
\end{center}
\caption{Exclusion boundaries from the conjecture $\EPNM>\EUG$.  The solid 
red curve shows the fiducial bound from Equation~(\ref{Eq.par0}) with the 
parameters of 
Equation~(\ref{Eq:PS1}); the shaded region is excluded.  The dotted curve shows the 
analytic bound from Equation~(\ref{Eq.sl}) and the dashed curve shows the bound as 
modified by Equations~(\ref{eq:kn-l}) and~(\ref{eq:qn-l}).  Triangles along the 
curves, from top to 
bottom, show the points where $u_t=[0.4,0.5,0.75,1.0,1.5,2.0,3.0]$ (along the 
solid and dashed curves the points begin with $u_t=0.5$, and along the 
solid (dashed) curve 
no points with $u_t>1.5 (2.0)$ are displayed).  For reference, symmetry 
parameters from the compilations of \cite{Dutra12,Dutra14} are indicated. The 
plus and open circle symbols (denoted `r') show interactions that were rejected, 
and the filled circles and open diamonds (denoted `a') show interactions that were 
accepted, as discussed later in Section~\ref{sec:rc}.}
\label{fig:sn-l}\end{figure}
%%%%%%%%%%%%%%%%%%%%%%%%%%%%%%%%%%%%

If we choose a value of $S_0$ that is greater than $S_0^\mathrm{LB}$, there 
exist two curves for $S(u)$ that can become tangent to $S^\mathrm{LB}(u)$: one 
at a density of $u_t>1$ and the other at a density of $u_t<1$. Each curve thereby satisfies
the inequality in Equation~\eqref{Eq:ineq}.  For every value of $S_0>S^\mathrm{LB}_0$,
there are thus two limiting values for $L$ that are each proportional to the 
slope of these tangential symmetry energy curves at $u=1$. The upper (or lower) bound 
for $L$ has a tangency density of $u_t<1$ (or $u_t>1$), as shown by the green 
lines in Figure~\ref{Fig:S_n}.\footnote{Note that for this example the lower bound 
for $L$ is negative.} The
conditions 
\begin{equation}
S(u_t)=\SLB(u_t)\qquad{\rm and}\qquad \left. {dS\over du}\right|_{u_t}=\left.{ d\SLB\over du}\right|_{u_t}
\label{eq:tan}\end{equation} 
at the tangency densities $u_t$ lead to the parametric equations
\begin{align}
S_{0}&=\frac{\EUG^0}{3u_t^{1/3}}(u_t+2)+\frac{\KPNM}{18}(u_t-1)^2+\frac{Q_n}{81}(u_t-1)^3-E_0\,,\nonumber\\
L&=\frac{2\EUG^0}{u_t^{1/3}}-\frac{\KPNM}{3}(u_t-1)-\frac{Q_n}{18}(u_t-1)^2.
%S_{0}&=\EUG^0\,u_t^{-1/3}(u_t+2)/3+\KPNM\,(u_t-1)^2/18-E_0,\nonumber\\
%L&=2\EUG^0\,u_t^{-1/3}-\KPNM(u_t-1)/3\,.
\label{Eq.par0}
\end{align}
When $u_t=1$, we recover the bounds $S_0=S_0^\mathrm{LB}$ and $L=L_0$.  
For every value of $u_t\ne1$, one can then determine a point of the boundary 
of the excluded region for $S_0(L)$, as shown in Figure~\ref{fig:sn-l}.
\footnote{By expanding $\EUG(u)$ around $u=1$ to the second order and dropping the skewness terms,
one can obtain an approximate but analytic form of this boundary as
$S_0=\EUG^0-E_0+(L-2\EUG^0)^2/[2(2\EUG^0+\KPNM)]$.}

The $(S_0,L)$ boundary depends on the saturation and UG parameters
$n_0, E_0, K_n, Q_n$, and $\xi_0$ and becomes less exclusionary the smaller 
$\EUG^0\propto n_0^{2/3}\xi_0$, or the larger $E_0$ and $\KPNM$, or the smaller 
(larger) $Q_n$ for $u_t<1$ ($u_t>1$). 

The Bertsch parameter for spin-half fermions is experimentally measured to 
be
$\xi_0=0.376 \pm 0.004$~\citep{Ku12}
or
$\xi_0=0.370 \pm 0.005 \pm 0.008$~\citep{Zurn13}.
The values of the saturation parameters are
$E_0=-15.9 \pm 0.4~\mev$,
$n_0=0.164\pm 0.007~\mathrm{fm}^{-3}$~\citep{Drischler16}, and
$K_0=240\pm 20~\mathrm{MeV}$~\citep{Shlomo06,Piekarewicz10}
or
$K_0=230\pm 40~\mathrm{MeV}$~\citep{Khan12}.
In general, $\Ksym<0$ for realistic relativistic mean field (RMF) and Skyrme 
forces, i.e., those that have been fit to properties of laboratory nuclei.  
Realistic microscopic interactions also suggest $\Ksym<0$: e.g., 
N$^3$LO chiral EFT calculations~\citep{Tews13} yield 
$\KPNM=119\pm101 \mev$~\citep{Margueron}.  The neutron-matter 
calculations of \cite{Drischler16} yield $-70$ MeV$>K_{\rm sym}>-240$ MeV 
and 10 MeV$<\KPNM<100$ MeV.
Since both experimental data and theoretical neutron-matter calculations 
indicate that $K_{\rm sym}<0$, it follows that $K_0>\KPNM>K_{n,0}$.   

Experimental constraints on $Q_0$ are weak. \cite{Cai14} have argued, 
based on heavy-ion flow data analyzed by \cite{Danielewicz02} and the
existence of $2M_\odot$ neutron stars, that -$494$ MeV$<Q_0<-10$ MeV. 
This range is consistent with that suggested by the neutron-matter 
calculations of \cite{Drischler16}, $-450$ MeV$<Q_0<-50$ MeV. Fitting 
the energies of the giant monopole resonance, \cite{Farine97} argue that 
$-1200$ MeV $<Q_0<-200$ MeV, which is also consistent with these 
other results. Constraints on $Q_{\rm sym}$ and $Q_n$ are even weaker 
than for $K_n$, but the neutron matter calculations of \cite{Drischler16} 
give $-750$ MeV$<Q_n<-250$ MeV.

From these parameter ranges and Equation~\eqref{Eq:ineq}, $\KPNM=K_0$ is 
a conservative (least exclusionary) choice for the bound. Since $(u_t-1)$ 
changes sign at $u_t=1$,  $Q_n=0$ is a conservative choice for $u_t>1$ 
while $Q_n=-750$ MeV is a conservative choice for $u_t<1$. Thus we adopt
\begin{align}\begin{split}
E_0&=-15.5~\mev,\
n_0=0.157~\mathrm{fm}^{-3},\
K_n=K_0=270~\mev,\\
\Ksym&=0,\
Q_n=0~\mev {\rm~or~}-750~\mev,\
\xi_0=0.365,
\label{Eq:PS1}
\end{split}\end{align}
as the fiducial conservative parameter set.
%based on experimental measurements, $1\sigma$ upper or lower bound
This results in 
$\EUG^0=12.64~\mev$, $S_0^\mathrm{LB}=28.14~\mev$,
$L_0=25.28~\mev$, and $K_{n,0}=-25.28~\mev$.

Figure~\ref{fig:sn-l} shows the bounds on $S_0$ and $L$ from the UG
constraint with the fiducial parameter set.  Points for several values of 
$u_t$ are shown, e.g., the point where $u_t=1/2$ is indicated, 
for which $L\simeq87$ MeV and $S_0\simeq33.6$ MeV. This value of $u_t$ 
could represent a plausible limit of applicability of the Taylor expansion of
Equation~(\ref{Eq:Sexp}). \footnote{Equation~(\ref{Eq:Sexp}) predicts $S(u\to0)=S_0
-L/3+\Ksym/18- Q_{\rm sym}/162$ whereas $S(u\to0)$ should, in fact, vanish.}  
From Equation~(\ref{Eq.par0}) one finds that $L=0$ MeV when $u_t=1.26$.  
Therefore, it appears that our bounds should be reliable for most of the 
figure since $u_t$ remains relatively close to unity throughout.
%%%%%%%%%%%%%%%%%%%%%%%%%%%%%%
%%%%%%%%%%%%%%%%%%%%%%%%%%%%%%
\begin{figure}[t]
\centerline{
\includegraphics[width=0.48\textwidth]{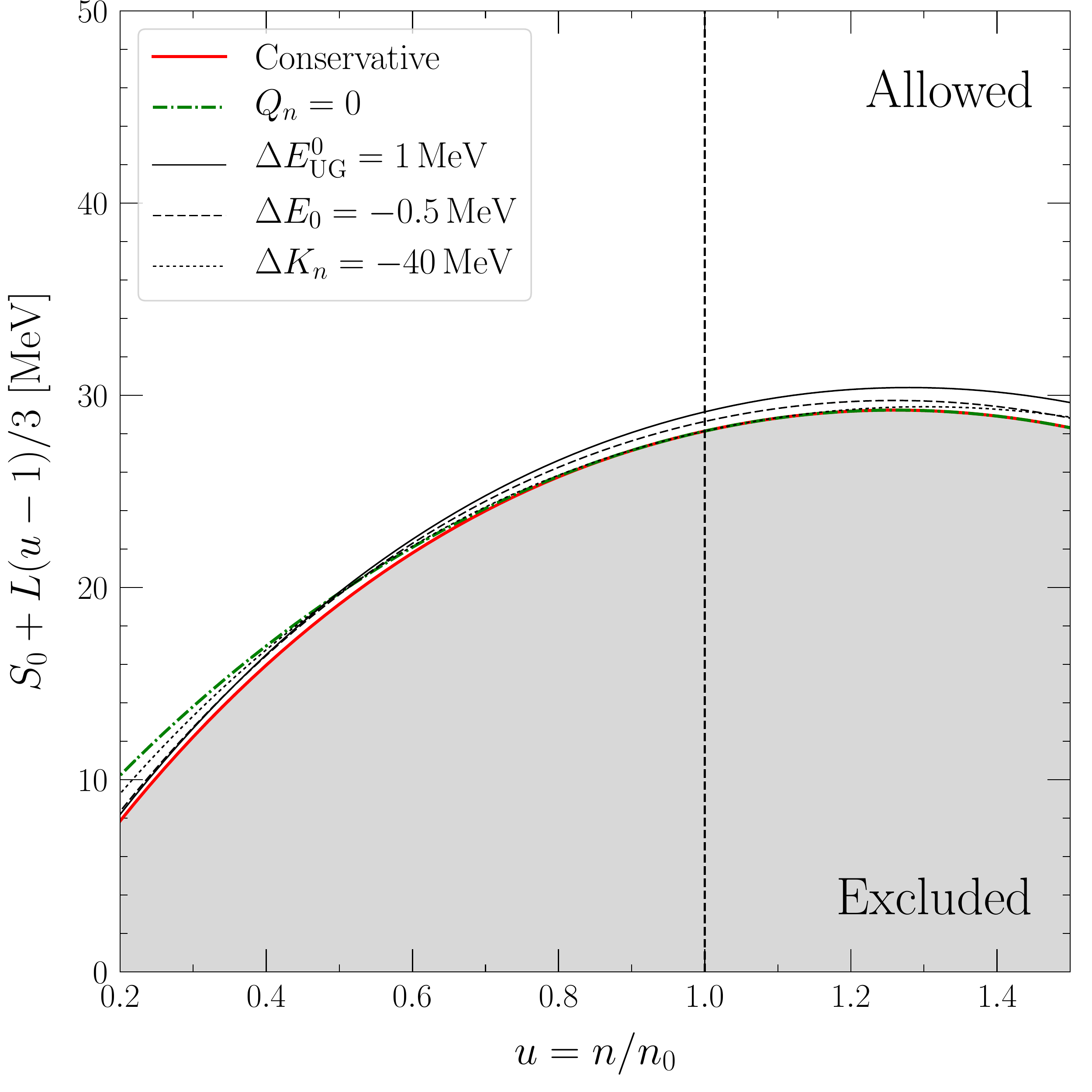}
\includegraphics[width=0.48\textwidth]{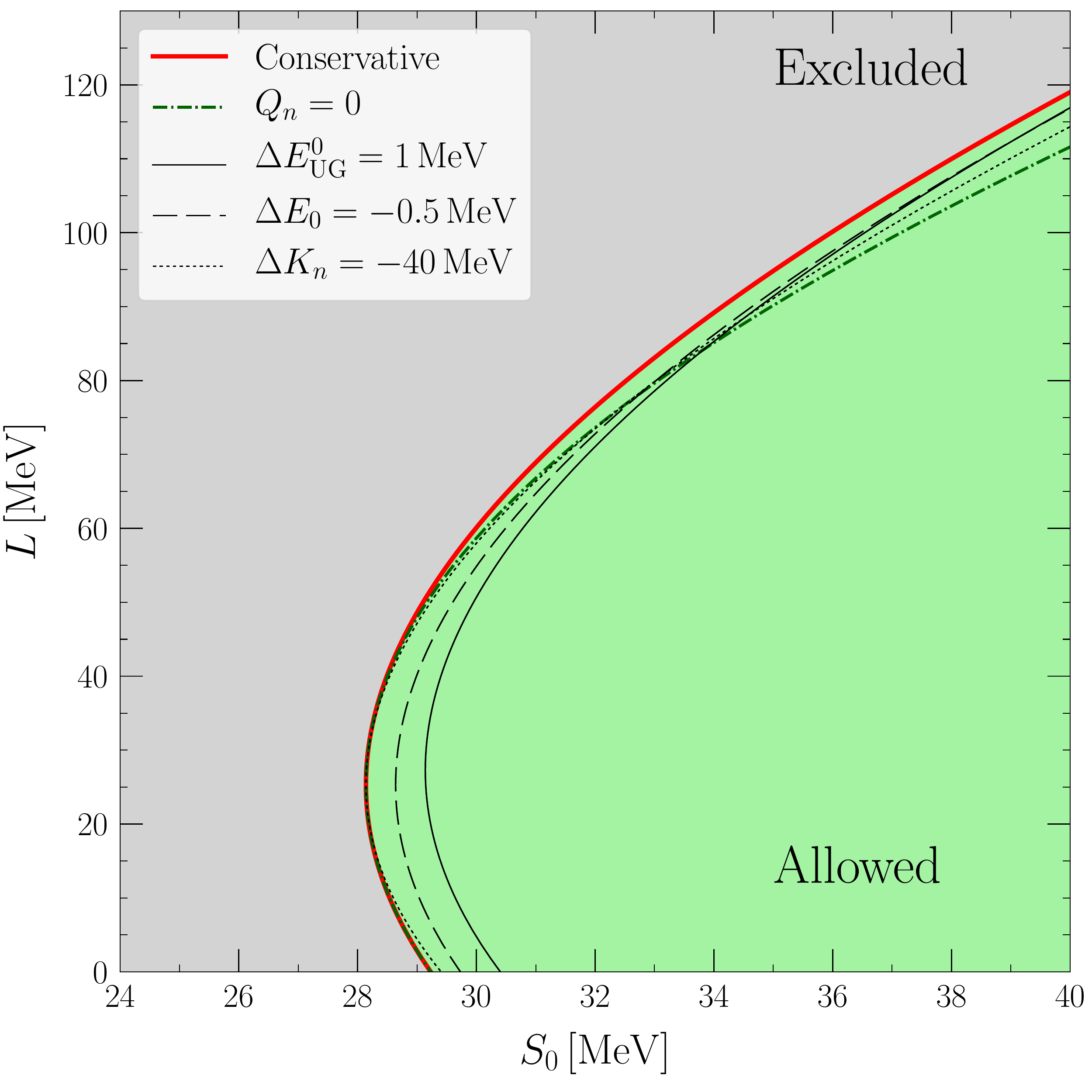}
}
\caption{Left panel: The parameter dependence of the bound for the
  symmetry energy.  Right panel:
  The parameter dependence of the bounds for the symmetry energy
  parameters $S_0$ and $L$.  In each panel, the solid curve shows the
  bound with the conservative parameter set, per Equation~\eqref{Eq:PS1}. 
Dotted-dashed, thin solid, dashed, and dotted lines show the
  bounds with $\Delta Q_n=750$ MeV (for $u_t<1$), $\Delta\EUG^0=1 \mev$, $\Delta E_0=-0.5 \mev$,
  and $\Delta K_n=-40 \mev$, respectively.}
\label{Fig:Sn_par}
\end{figure}
%%%%%%%%%%%%%%%%%%%%%%%%%%%%%%

Figure~\ref{Fig:Sn_par} shows the effect of changing assumptions about
the saturation and unitarity parameters. Note that only variations in
$E_0$, $K_n, Q_n$, and $\EUG^0$ need to be considered.  Results are
shown for the variations $\Delta \EUG^0=~+1~\mev$, $\Delta
E_0=-0.5~\mev$, and $\Delta K_n=-40~\mev$, as well as for $\Delta
Q_n=+750$ MeV when $u_t<1$.  Variations caused by $|\Delta
Q_n|\sim300$ MeV are similar to those from changing $K_n$ by
about 30 MeV.  Varying $E_0$ or $\EUG^0$ translates directly into a
movement of the point $(S_0^\mathrm{UB},L_0)$ and thus produces 
the largest variations in the exclusion boundary near $u_t\sim1$.  In this
region, variations within the large uncertainties for $K_n$ and $Q_n$ are 
negligible. These uncertainties only become significant when
$u_t$ is appreciably different from unity, where the validity of our
bounds is not assured.  Nevertheless, variations of $K_n$ 
of the order of 30 MeV and $Q_n$ of the order of 300 MeV translate into
boundary shifts of no more than $\sim1$ MeV in $S_0$ for a given value
of $L$, or $\sim3$ MeV in $L$ for a given value of $S_0$, even for the
extreme case $|u_t-1|\simeq0.8$.  We conclude that the fiducial
boundary is remarkably insensitive to uncertainties in the saturation
parameters and the Bertsch parameter.

\cite{Zhang17} recently argued that the 
symmetry energy bounds are much more uncertain than we have 
suggested because the uncertainties in $\xi_0$, $K_0, K_{\rm sym}$, 
$Q_0$, and $Q_{\rm  sym}$ are much greater than proposed here.  For 
example, \cite{Zhang17} argued that the uncertainty in $\xi_0$ is 
about 0.1, but this is not supported by recent cold atom experiments, as
discussed in this section.  Furthermore, we have found that the individual 
parameters $K_0, K_{\rm sym}, Q_0$, and $Q_{\rm sym}$ themselves are 
not particularly relevant for the symmetry bounds. Rather, it is the 
combinations $K_n=K_0+K_{\rm sym}$ and $Q_n=Q_0+Q_{\rm sym}$ that 
directly appear.  Moreover, the bulk of nuclear interactions developed over 
decades to fit nuclear properties show pronounced correlations among 
$K_n, Q_n$, and $L$.  As a result, the large individual parameter uncertainties 
do not play as important a role as may have been thought. A more realistic 
symmetry bound incorporating these correlations is addressed in the next
section.

\section{More Realistic Constraints on the Symmetry Energy Parameters\label{sec:rc}}
The previous section described a minimal constraint for the 
symmetry energy parameters as obtained from relatively conservative 
choices for the saturation properties of matter and the Bertsch parameter.
The largest parameter uncertainties exist for $K_{\rm sym}$, $Q_0$, and
$Q_{\rm sym}$.  In this section, we discuss the evidence that these 
parameters are correlated with each other and with $L$.  Such collective 
correlations can reduce the variations introduced by the uncertainties of 
the individual parameters. The use of these correlations results in a more 
phenomenological but possibly more realistic symmetry energy bound. 

\begin{figure}[t]
\begin{center}
\includegraphics[width=.6\textwidth]{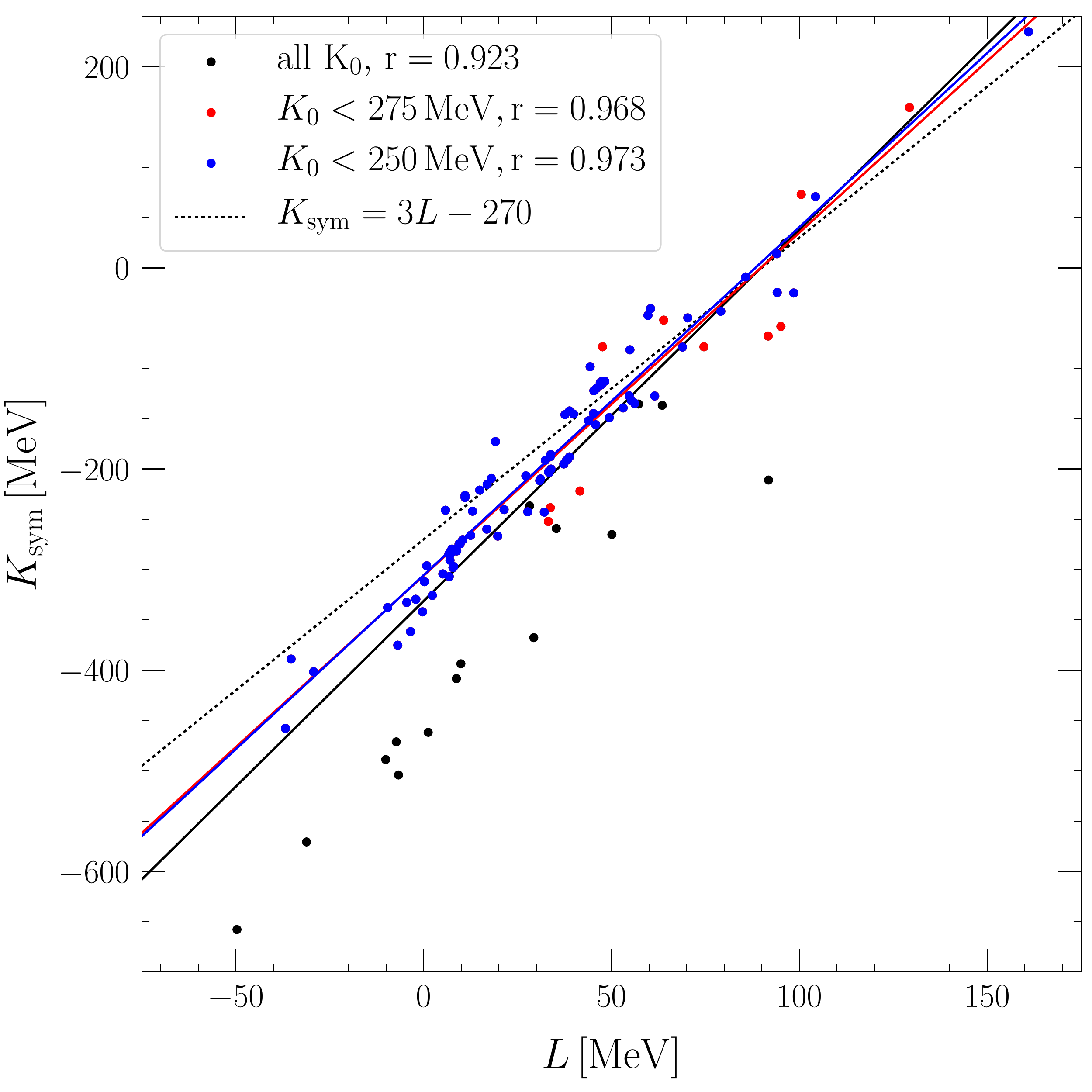}
\end{center}
\caption{Correlations between $K_{\rm sym}$ and $L$ for 118 Skyrme
  forces studied by \cite{Danielewicz09}.  Colors distinguish forces
  according to their the incompressibility parameters $K_0$. Their respective
 correlation coefficients are given by $r$.}
\label{fig:ksym}
\end{figure}

In our fiducial estimate, we conservatively chose $K_\mathrm{sym}=0$.
More realistically, however, $\Ksym<0$. It has also been found that realistic 
forces of both nonrelativistic potential and relativistic field-theoretical types 
that are calibrated by fitting energies and charge radii of laboratory nuclei 
show a linear correlation between $\Ksym/S_0$ and 
$L/S_0$~\citep{Danielewicz09} or between $\Ksym$ and
$L$~\citep{Chen09,Vidana11,Ducoin11}.\footnote{Since the variation of
$S_0$ among different forces is relatively small, these correlations are 
essentially the same.}  Specifically, \cite{Danielewicz09}
studied 118 Skyrme forces and found the correlations to be bracketed by 
the expressions
\begin{equation}
\Ksym/S_0=6L/S_0-18\qquad{\rm~and~}\qquad\Ksym/S_0=4L/S_0-10\,,
\label{eq:DanLee}\end{equation}
while \cite{Chen09} studied 63 Skyrme interactions and the MDI force
and found the bracketing expressions
\begin{equation}
\Ksym=5L-500~\mev\qquad{\rm~and~}\qquad\Ksym=4L-250~\mev\,.
\label{eq:Chen}\end{equation}

However, studying the same interactions as \cite{Danielewicz09}, we have 
found the correlation
\begin{equation}\Ksym\simeq3.69L-331.2\pm41.5~\mev,\quad\qquad r=0.923\,.
\label{Eq.ksymall}
\end{equation}
The poor quality of this correlation is largely due to the wide range of $K_0$ 
values among the forces.  As observed in Figure~\ref{fig:ksym}, when interactions 
with $K_0>275\,(250) ~\mev$ are excluded from consideration, 
the correlation is considerably tightened.  We have found
\begin{align}
\Ksym&\simeq3.41L-306.0\pm28.3~\mev,\quad r=0.968\quad(K_0<275 \mev),\\
\Ksym&\simeq 3.46L-305.5\pm26.4~\mev,\quad r=0.973\quad(K_0<250 \mev),
\label{Eq.ksym}
\end{align}
where the uncertainty is such that 68.3\% of the interactions with
$K_0\le275\,(250) ~\mev$ lie within these bounds.  We verified that similar
correlations also exist between $\Ksym/S_0$ and $L/S_0$, which are slightly
more significant.

The facts that the intercept of this correlation is close to our
upper limit to $K_0$ and that the slope is close to 3 fortuitously allow the
simplification $\KPNM \simeq 3L$.  With this expression, the boundary
in Equation~\eqref{Eq.par0} can be analytically expressed (still assuming $Q_n=0$)
as
\begin{align}
S_{0}&=\frac{\EUG^0}{3}\frac{1+2u_t^2}{u_t^{4/3}}-E_0
\ ,\quad
L=\frac{2\EUG^0}{u_t^{4/3}}\,.
\label{Eq.par2}
\end{align}
Eliminating $u_t$, one finds, simply,
\begin{align}
%S_{0}=L\,\left[1+2(2\EUG^0/L)^{3/2}\right]/6-E_0\ .
S_{0}=\frac{L}{6}\,\left[1+2\left(\frac{2\EUG^0}{L}\right)^{3/2}\right]-E_0\ .
\label{Eq.sl}
\end{align}
This analytic exclusion boundary is compared to the fiducial conservative,
boundary in Figure~\ref{fig:sn-l}.  We note that the analytic exclusion boundary 
does not allow negative values of $L$.

%%%%%%%%%%%%%%%%%%%%%%%%%%%%%%%%%%%%%%%
\begin{figure}[t]
\includegraphics[width=.48\textwidth]{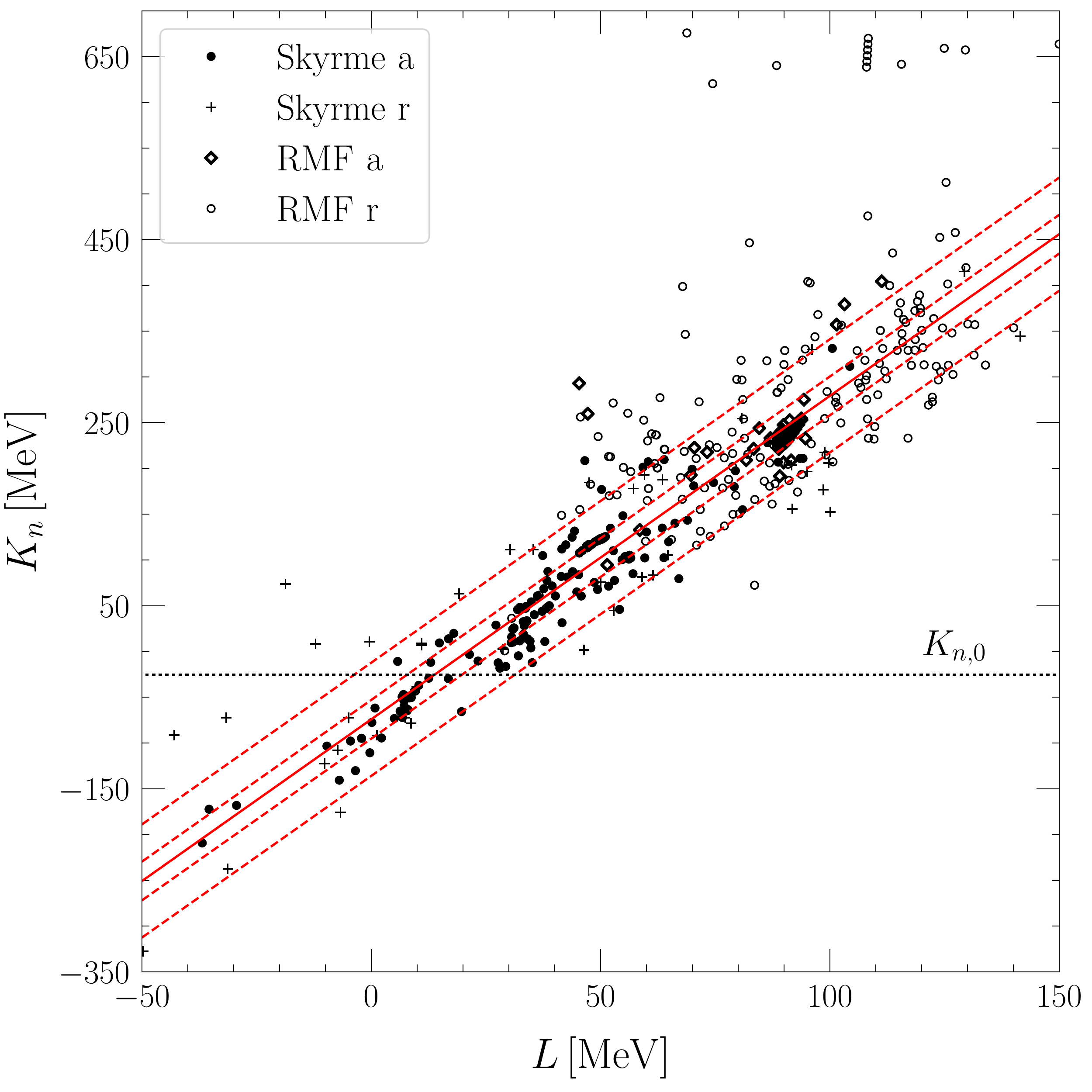}
\includegraphics[width=.48\textwidth]{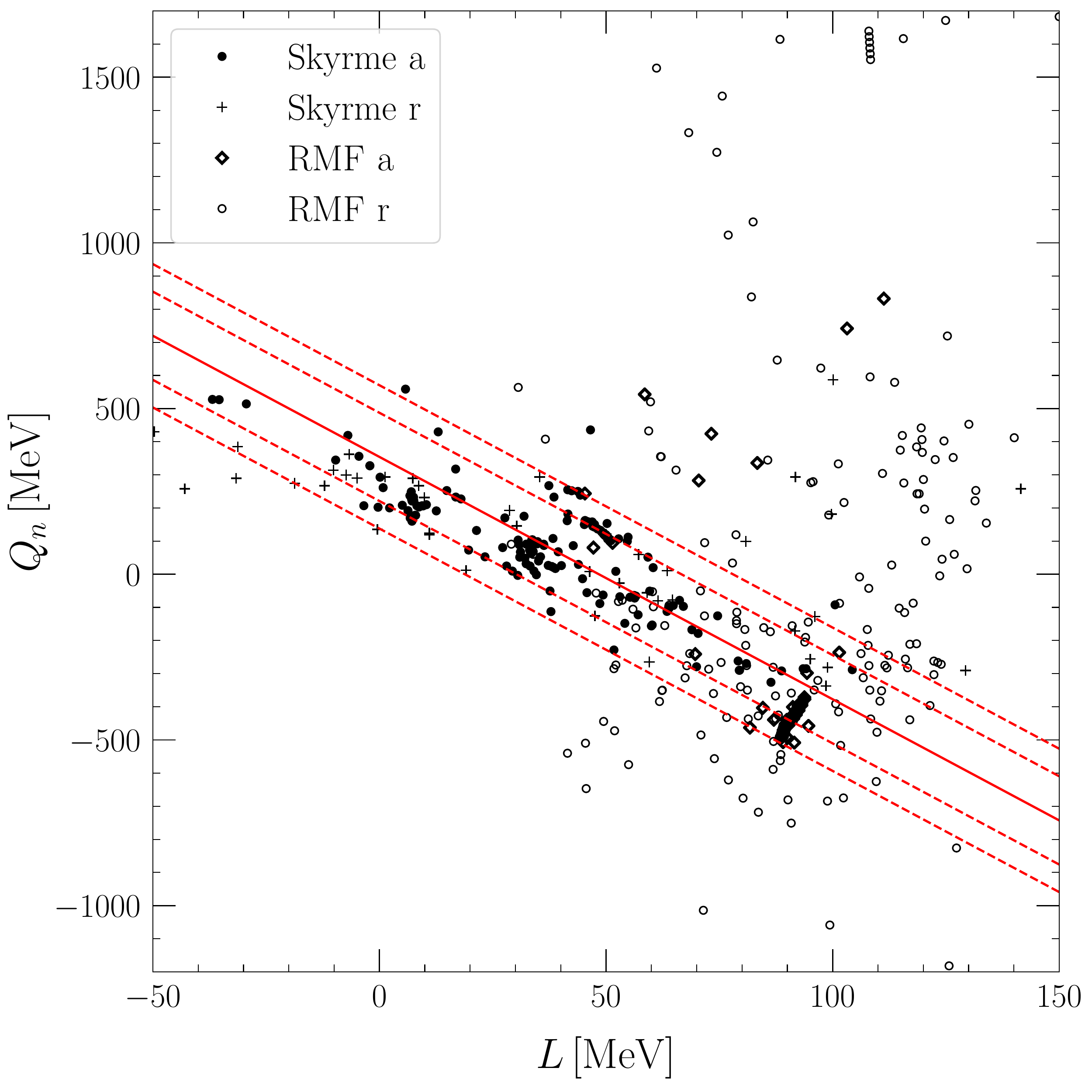}
\caption{The left (right) panel shows the neutron-matter incompressibility $K_n$ (skewness $Q_n$) versus the symmetry parameter $L$ for 240 Skyrme interactions compiled by \cite{Dutra12} and 263 RMF forces compiled by \cite{Dutra14}. The solid lines show the correlations obtained from the `accepted' (a) realistic interactions after excluding the `rejected' (r) interactions; the dashed lines enclose, respectively, 68.3\% and 95.4\% of the accepted interactions.  The dotted line in the left panel shows the minimum value -25.28 MeV for $K_n$ permitted by the conjecture that $\EPNM\ge\EUG$.}
\label{fig:kn-l}\end{figure}
%%%%%%%%%%%%%%%%%%%%%%%%%%%%%%%%%%%%%%%%%%

\begin{figure}[t]
\includegraphics[width=.48\textwidth]{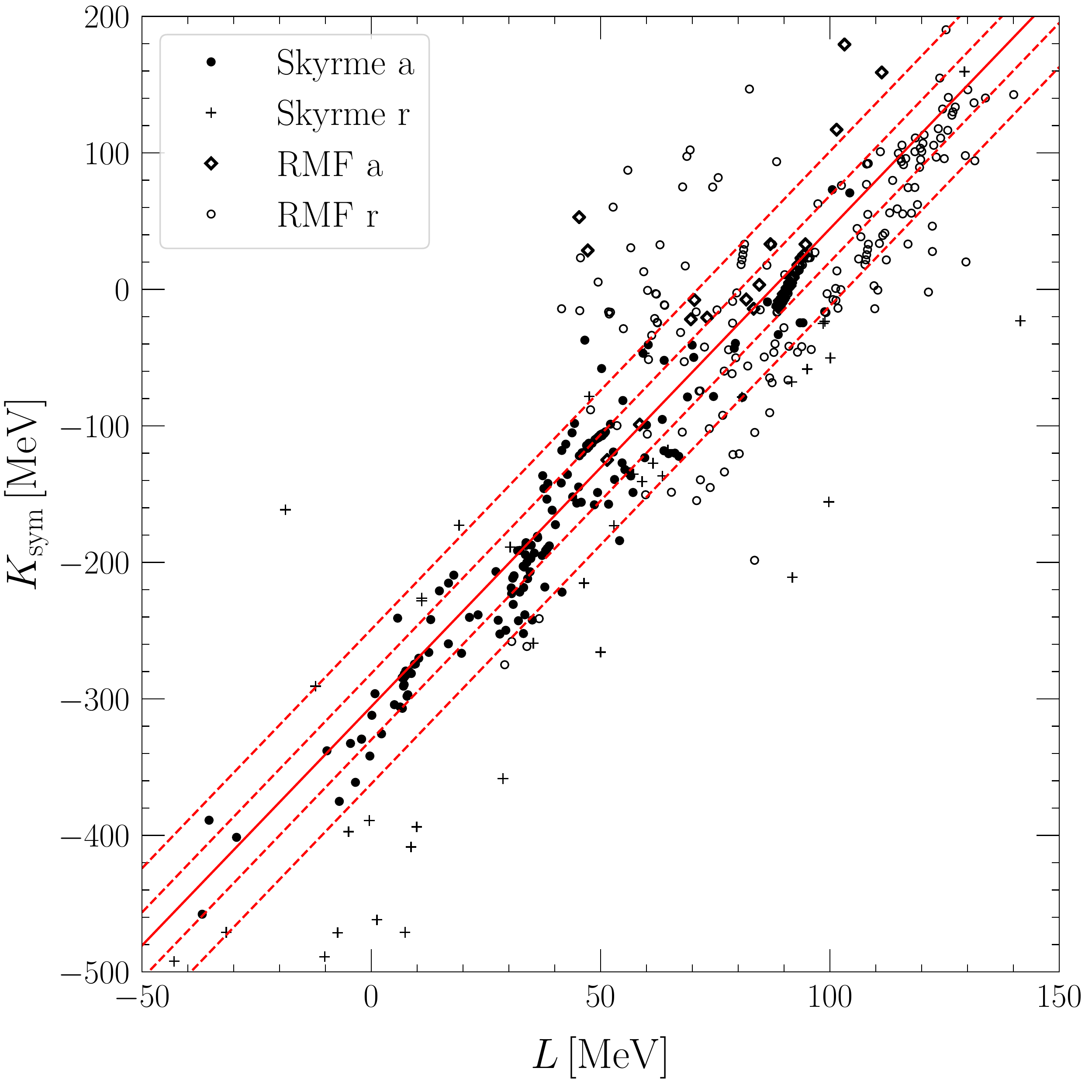}
\includegraphics[width=.48\textwidth]{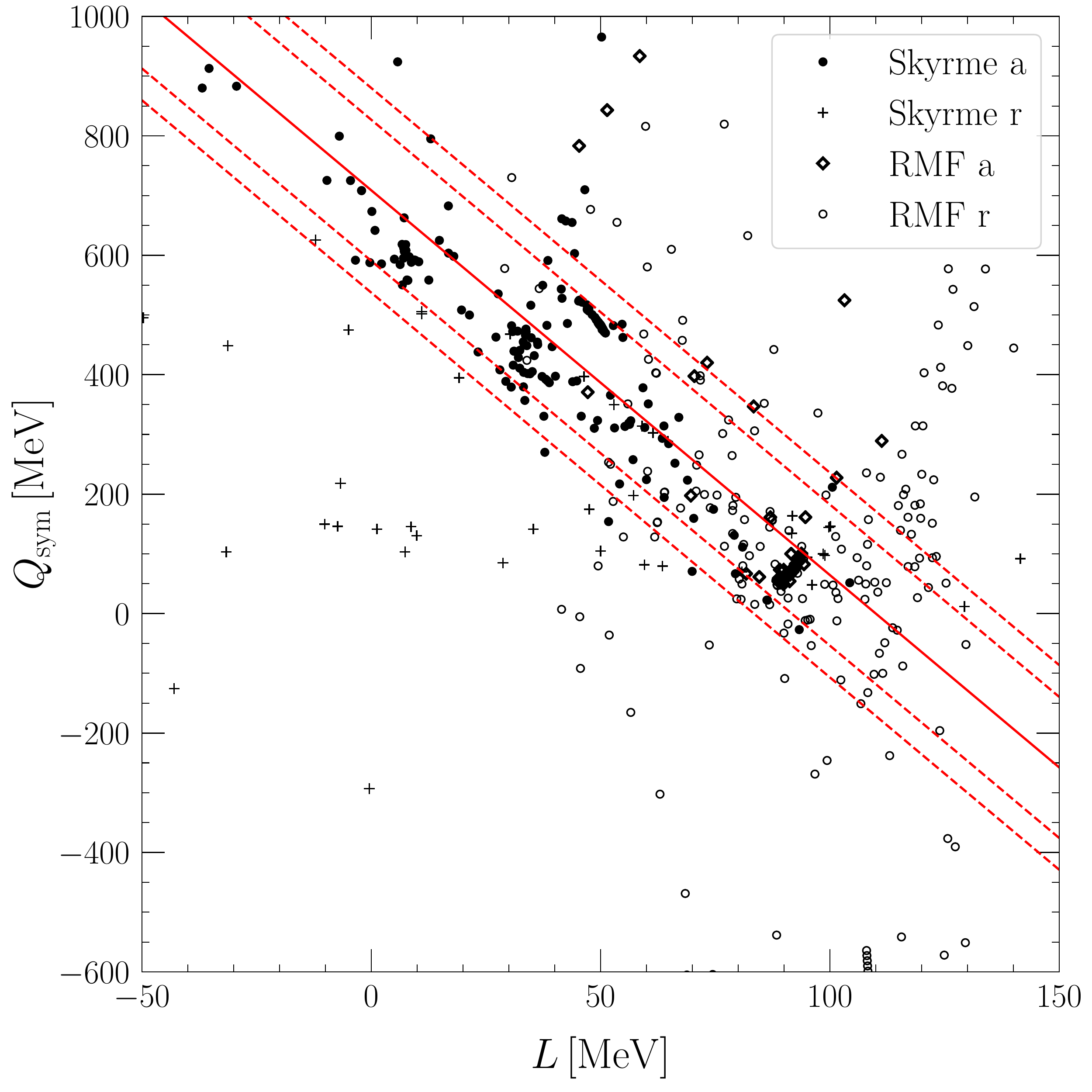}
\caption{The left (right) panel shows the symmetry incompressibility $K_{\rm sym}$ (skewness $Q_{\rm sym}$) versus the symmetry parameter $L$ for 240 Skyrme interactions compiled by \cite{Dutra12} and 263 RMF forces compiled by \cite{Dutra14}.   The solid lines show the correlations obtained from the `accepted' (a) realistic interactions after excluding the `rejected' (r) interactions; the dashed lines enclose, respectively, 68.3\% and 95.4\% of the accepted interactions.}
\label{fig:ksym-l}\end{figure}

\begin{figure}[t]
\begin{center}
\includegraphics[width=.6\textwidth,angle=0]{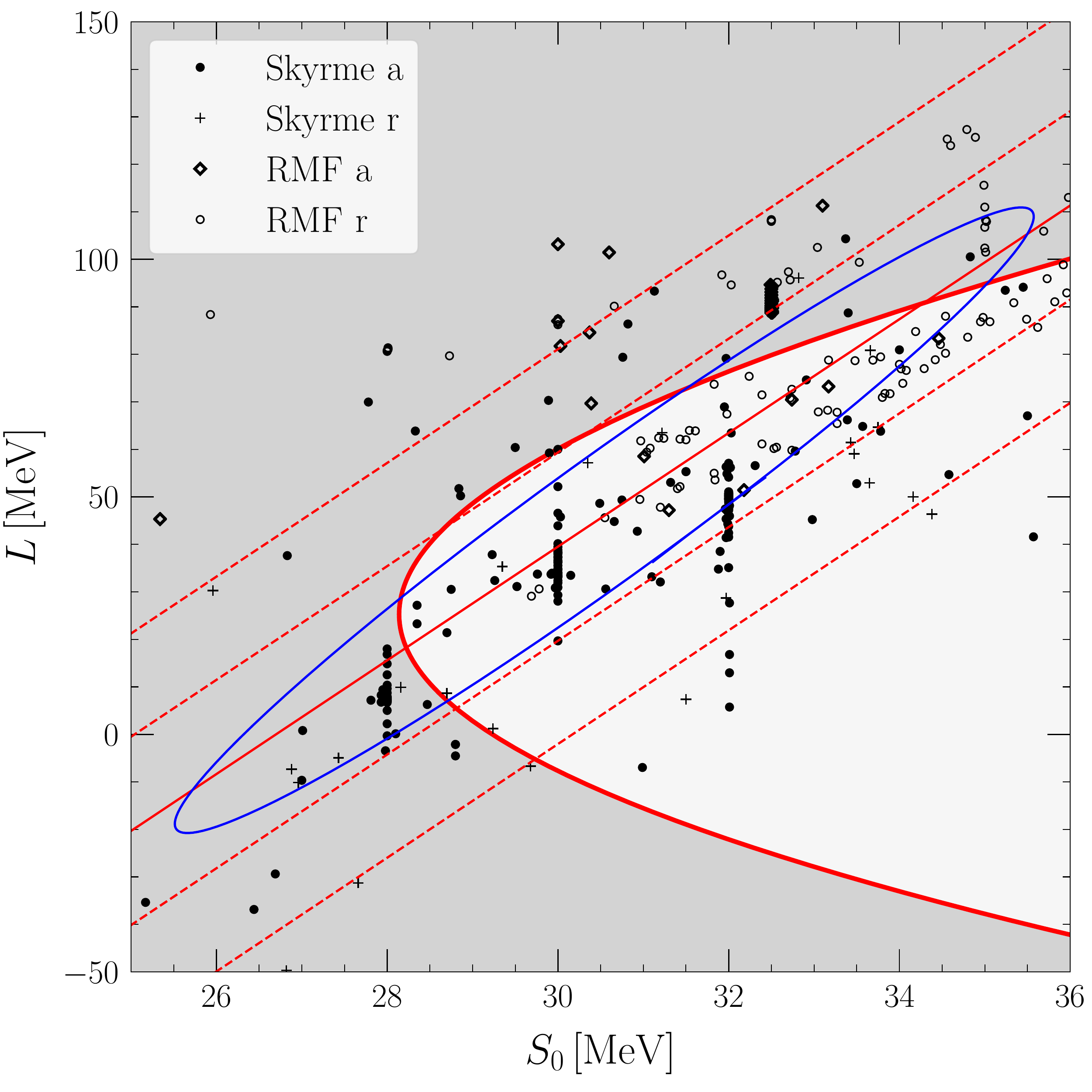}
\end{center}
\caption{The confidence ellipse for the $S_0-L$ correlation as
  determined by \cite{Kortelainen10} for an assumed fiducial binding 
  energy error of 2~MeV.  The solid red curve shows the conservative 
  fiducial exclusion boundary from Equation~(\ref{Eq.par0}) with the parameter 
  set from Equation~(\ref{Eq:PS1}). For reference, the symmetry parameters from 
  the compilations of \cite{Dutra12,Dutra14} are indicated, as in 
  Figure~\ref{fig:kn-l}. The correlation between these parameters for the
  `accepted' interactions of Equation~\eqref{eq:S0-l} is shown as the solid line, 
  together with the dashed lines showing 68.3\% and 95.4\% enclosures.}
\label{fig:sn-l-be}\end{figure}

We now examine correlations among the parameters in more detail. 
A collection of 240 Skyrme forces was studied by \cite{Dutra12} and another 
collection of 263 RMF forces by \cite{Dutra14}. Using 
these compilations, we have found that $K_n$, $Q_n$, and $L$ are strongly
correlated. However, these compilations, especially the one containing RMF 
interactions, include many parametrizations with very unrealistic saturation 
properties. Keeping the 188 Skyrme and 73 RMF interactions
with 0.149~fm$^{-3}<\rho_0<0.17$ fm$^{-3}$, $-17~\mev<E_0<-15~\mev$,
$25~\mev<S_0<36~\mev$, and $180~\mev<K_0<275~\mev$ (accepted 
interactions) and rejecting the others, we find the following correlations with 
the respective correlation coefficients $r$:
\begin{align}
K_n&=3.534~L-(74.02\pm21.17~[61.84])~\mev,\quad\qquad &r=0.96\,,\label{eq:kn-l}
\end{align}
and
\begin{align} 
Q_n&=-7.313~L+(354.03\pm133.16~[216.3])~\mev,\qquad &r=-0.78\,.\label{eq:qn-l}
\end{align}
The quoted errors correspond to displacements in a correlation line that will 
enclose 68.3\% of the interactions (95.4\% for the error displayed in square 
brackets). We show correlations in Figure~\ref{fig:kn-l}. In this figure, as well as 
in Figures~\ref{fig:sn-l},~\ref{fig:ksym-l}, and~\ref{fig:sn-l-be}, many of the `rejected' 
interactions fall outside the boundaries of the plot. 

Note that $K_n$ and $L$ are especially highly correlated ($r=0.96$).  
Furthermore, these results indicate that $K_n$ and $Q_n$ are negatively 
correlated while their errors are positively correlated: an interaction that has 
a $K_n$ value greater than the mean for its $L$ value also tends to have 
a $Q_n$ value larger than the mean. In the left panel of Figure~\ref{fig:kn-l}, we 
also indicate the minimum value of $K_n$ permitted by the conjecture that
$\EPNM\ge\EUG$, namely $K_{n,0}=-25.28~\mev$. Using the correlation
of Equation~(\ref{eq:kn-l}), this implies a lower limit to $L$, $-3.7~\mev$.
%%%%%%%%%%%%%%%%%%%%%%%%%%%%%%%%%%%%%%%
%%%%%%%%%%%%%%%%%%%%%%%%%%%%%%%%%%%%%%%%%%

In a similar way, the parameters $S_0, K_{\rm sym}, Q_{\rm sym}$, and $L$ are 
also strongly correlated:
\begin{align}
L &=11.969 S_0 - (319.55 \pm 19.83 [41.56])~\mev, \qquad &  r= 0.74\,,
\label{eq:S0-l}\\
K_{\rm sym}&=3.501~L-(305.67\pm24.26~[56.59])~\mev,\qquad & r=0.96\,,\label{eq:ksym-l}
\end{align}
and 
\begin{align}
Q_{\rm sym}&=-6.443~L+ (708.74\pm118.14~[171.34])~\mev,\qquad &r=-0.86\,,\label{eq:qsym-l}
\end{align}
as depicted in Figures~\ref{fig:ksym-l} and~\ref{fig:sn-l-be}.  Note that the correlation
between $K_{\rm sym}$ and $L$ is not surprisingly nearly the same as
that found from the compilation of \cite{Danielewicz09},
Equation~(\ref{Eq.ksym}).  As for $K_n$ and $Q_n$, there is a negative
correlation between $K_{\rm sym}$ and $Q_{\rm sym}$, as well as a
positive correlation in their errors.

Note that correlations among the parameters are not as
strong for RMF approaches as they are for the nonrelativistic Skyrme models.
It is also noteworthy that a large fraction of these Skyrme interactions (22\%) 
but an even larger fraction of the RMF forces (73\%) have 
unrealistic saturation properties as established from fitting nuclear 
masses, e.g., using liquid-droplet mass formula fits (and we have, hence, 
`rejected' them). Therefore, such compilations should be used with caution.
\cite{Zhang17} used the same compilations to argue that uncertainties
in $K_n$ and $Q_n$ are too large to definitely constrain the
parameters $S_0$ and $L$ from UG arguments, but they did not
reject unrealistic interactions.  Curiously, the correlation between
$K_n$ and $Q_n$ is apparent in their study even without any
rejections, although it was not utilized.

Finally, a more realistic exclusion boundary is obtained by substituting
Equations~(\ref{eq:kn-l}) and (\ref{eq:qn-l}), both with their upper 95.4\%
error bars, into Equation~(\ref{Eq.par0}).  This allows the elimination of
$K_n$ and $Q_n$, so that once again $S_0$ and $L$ are solely related
through the parametric variable $u_t$.  This modified boundary is also
shown in Figure~\ref{fig:sn-l} in comparison to the fiducial boundary
from Equation~(\ref{Eq.par0}) and the analytic boundary given by 
Equation~(\ref{Eq.sl}).  
The modified boundary closely follows the analytic boundary.  We emphasize
that the errors of the correlations between $K_n$ and $L$ and between 
$Q_n$ and $L$ are small enough that these boundaries are affected 
at the 1~MeV level only for $|u_t-1|\simge0.2$, at which point the Taylor 
expansion of the energies is probably unreliable. Thus the analytic boundary in
Equation~(\ref{Eq.sl}) is a good approximation for the realistic boundary provided
by UG constraints plus correlations among $K_n$, $Q_n$, and $L$.

%%%%%%%%%%%%%%%%%%%%%%%%%%%%%%%%%%%
%%%%%%%%%%%%%%%%%%%%%%%%%%%%%%%%%%%%%%%%%%%%%%%%%

\section{Comparison with Experimental Constraints}
The most abundant, accurate, and model-free experimental constraint on
symmetry energy parameters comes from nuclear binding energies.  As
discussed by \cite{Lattimer13}, the resulting correlation between
$S_0$ and $L$ is nearly linear.  A recent Hartree-Fock study by the
UNEDF collaboration \citep{Kortelainen10} found the confidence ellipse
shown in Figure~\ref{fig:sn-l-be}, assuming a fiducial fitting error of 2~MeV for 
the nuclear binding energies.  A nearly identical result was found from a 
liquid-droplet analysis~\citep{Lattimer13}.  This confidence ellipse is 
essentially compatible with the UG constraint (for positive values 
of $L$).  Even though a large fraction of the parameter sets taken from the 
compilations of \cite{Dutra12,Dutra14} do not satisfy the UG 
constraint, a larger fraction of the accepted parameter sets do obey this 
constraint; furthermore, they show almost exactly the same correlation 
as in the analysis of \cite{Kortelainen10}.  Combined, the UG and 
nuclear binding-energy constraints imply that $L<100$ MeV and 
$S_0<35$ MeV.
%%%%%%%%%%%%%%%%%%%%%%%%%%%%%%%%%%
\begin{figure}[h]%[thbp]
\centering
\includegraphics[width=0.48\textwidth]{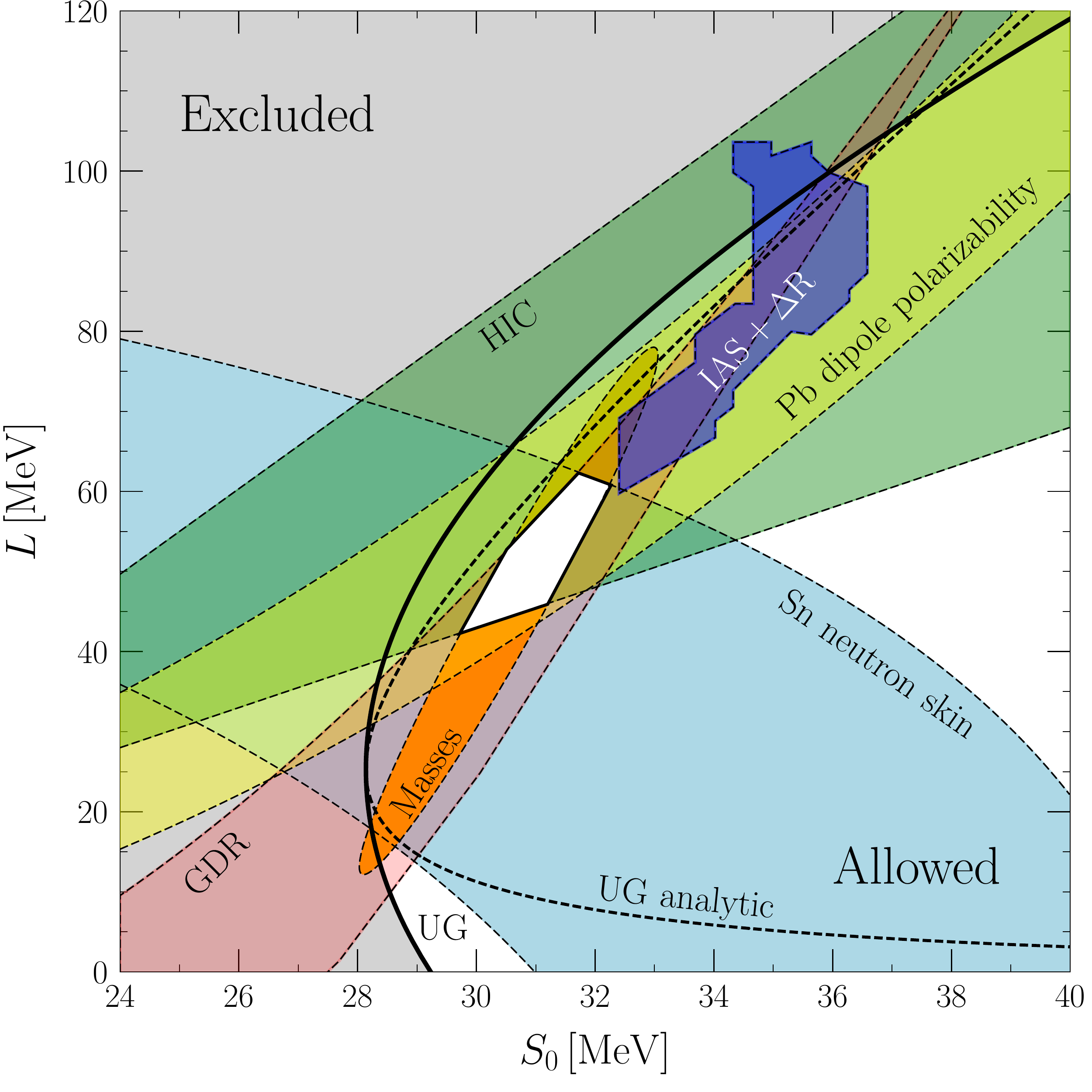}
\includegraphics[width=0.48\textwidth]{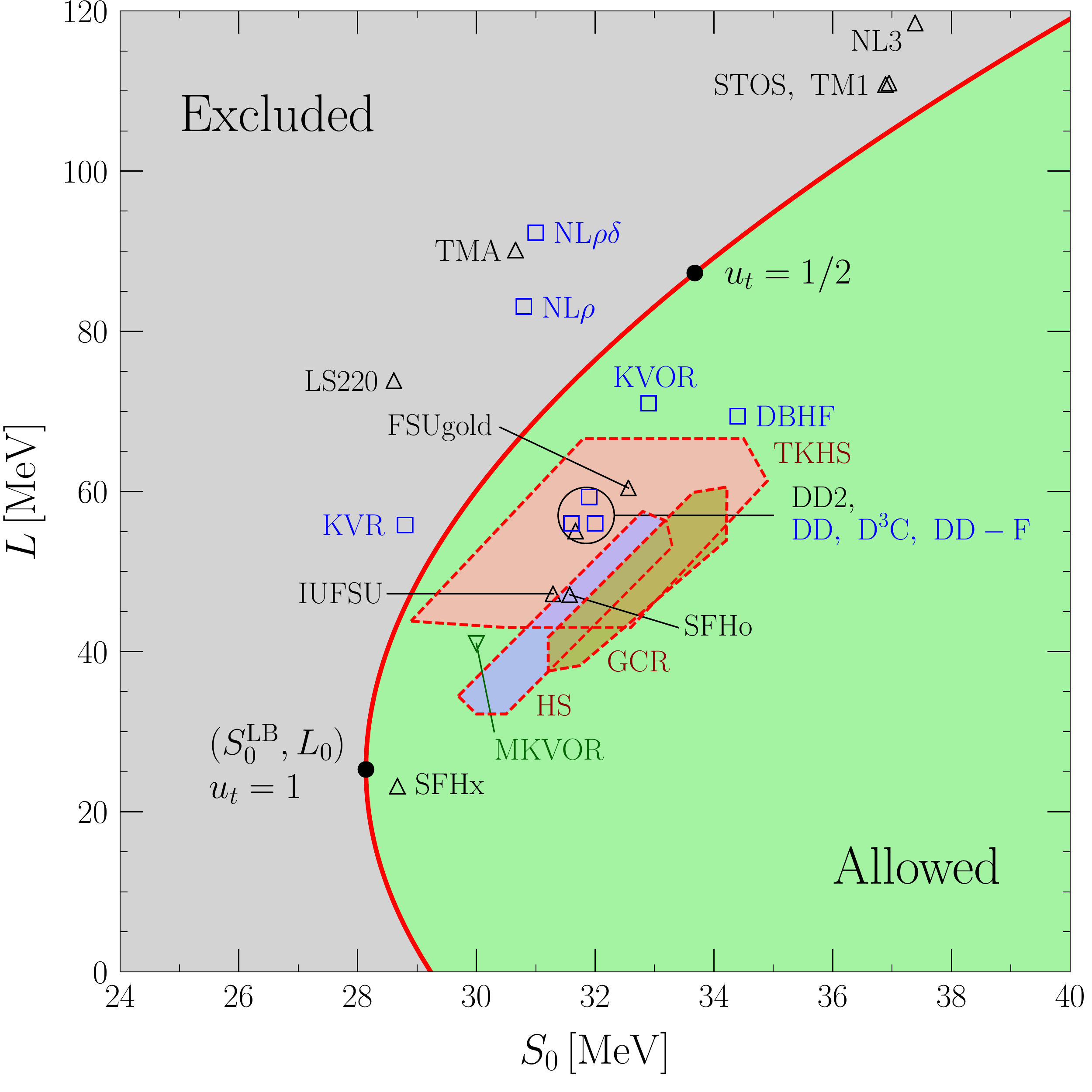}
\caption{UG bounds on symmetry energy parameters. The thick
  lines show the bound Equation~(\ref{Eq.par0}) using the conservative
  parameter set of Equation~(\ref{Eq:PS1}). Excluded regions are shown by shading.  
  Left panel: Experimental constraints are from \cite{Lattimer13} and 
  \cite{Lattimer14}, supplemented by isobaric analog states and isovector 
  skin (IAS+$\Delta R$) results from \cite{Danielewicz17}.  The thick dashed 
  curve shows the analytic bound from Equation~(\ref{Eq.sl}).  
  Right panel: Filled circles show the point $(S_0^\mathrm{LB},L_0)$ at the 
  tangent density $u_t=1$ and the point where $u_t=1/2$. Triangles show 
  values for  interactions commonly used in tabulated equations of state for
  astrophysical simulations (notation and data from \cite{Fischer14}), and 
  open squares (from \cite{Klahn07}) and 
  the inverted triangle (from \cite{MKV}) show those of other frequently 
  used interactions. The shaded regions TKHS, GCR, and HS
  show the parameter ranges inferred from the PNM calculations of
  \cite{Tews13}, \cite{Gandolfi12}, and \cite{HLPS:2010}, respectively.}
\label{Fig.sl1}
\end{figure}
%%%%%%%%%%%%%%%%%%%%%%%%%%%%%%%%%%%%%%%%%%%

Other experimental constraints on symmetry energy parameters are
reviewed in \cite{Lattimer13} and \cite{Oertel17}.  They indicate that
consistency with measurements of nuclear masses, giant dipole
resonances and dipole polarizabilities, neutron-skin thicknesses, and 
flows in heavy-ion collisions is achieved for 30 MeV~$\le S_0\le$~32 MeV 
and 40~MeV $\le L\le$~60 MeV (left panel of Figure~\ref{Fig.sl1}).\footnote{In 
the left panel of Figure~\ref{Fig.sl1}, the results of \cite{Kortelainen10} are 
displayed using a fiducial fitting error of 1 MeV, as using 2 MeV leads to 
negative $L$ values for small $S_0$.}  It is observed that these ranges 
for $S_0$ and $L$ are compatible with neutron-matter calculations 
and both conservative and realistic UG bounds. Recently, 
\cite{Danielewicz17} have argued, using isobaric analog states and isovector 
skins on neutron-rich nuclei, that both symmetry parameters may be larger 
than this consensus. Their 68\% confidence region, depicted in the left panel 
of Figure~\ref{Fig.sl1}, does not coincide with the overlap region from
\cite{Lattimer13} but is still mostly compatible with our UG constraint.  
This figure emphasizes the importance of both theoretical neutron-matter 
calculations, suitably calibrated by the energies of light nuclei, and
measurements of neutron-skin thicknesses in establishing an upper
limit to the parameter $L$.

\section{Applications\label{Sec:App}}

In the following we demonstrate the significance of the UG 
bound for astrophysics. In the right panel of Figure~\ref{Fig.sl1} we plot  
the values of the symmetry energy parameters for 10 tabulated equations of 
state \citep{Fischer14} frequently used for astrophysical simulations. Note that 
half of them violate the bound, which highlights the need for additional equation 
of state tables that satisfy these conservative constraints. We also display the 
results for other interactions~\citep{Klahn07,MKV} commonly used in astrophysics 
and heavy-ion physics, among which a nontrivial number are found to violate the 
bound.  We emphasize that realistic uncertainties in the relevant parameters 
$\xi_0, n_0, E_0, K_n$, and $Q_n$ do not affect these conclusions in any 
significant fashion.

Furthermore, the lower limit on the symmetry energy, implied by the UG 
constraint for
$u<1$, has implications for the surface energy of nuclei, the location of the 
crust-core boundary, and the radii and moments of inertia of neutron stars.  
We show herein that this lower limit will establish maxima to the surface 
symmetry energy parameter $S_S$ and minima to neutron-star radii and 
moments of inertia.  Curiously, although our conjecture $\EPNM>\EUG$ 
essentially determines a minimum for the symmetry energy, it 
also implies a maximum limiting behavior for $u\ge1$. This has 
implications for the threshold density for the onset of rapid neutrino cooling 
due to the nucleon Urca process and, thus, for neutron-star cooling.

To investigate these applications of the UG bound, we require a better 
parameterization of $S$ than that given by the expansion of Equation~(\ref{Eq:Sexp}), 
which fails in the limits of both small and large $u$. Instead, we model the 
symmetric matter and symmetry energy using these expressions:
\begin{align}
E_{\rm SNM}&=T\left[u^{2/3}+a^\prime u+b^\prime u^{4/3}+c^\prime u^{5/3}+d^\prime u^2\right],\label{eq:epar}
\end{align}
and
\begin{align}
S(u)&=T\left[\left(2^{2/3}-1\right)u^{2/3}+au+bu^{4/3}+cu^{5/3}+du^2\right].\label{eq:spar}
\end{align}
The parameters are fit to properties of matter at saturation density ($u=1$):
\begin{align}\begin{split}
a^\prime&=-4+{120E_0+6K_0-Q_0\over6T}\,,\\
b^\prime&=6+{-90E_0-5K_0+Q_0\over2T}\,,\\
c^\prime&=-4+{72 E_0+4K_0-Q_0\over2T}\,,\\
d^\prime&=1+{-60E_0-3K_0+Q_0\over6T}\,,\\
a&=-4\left(2^{2/3}-1\right)+{120S_0-38L+6K_{\rm sym}-Q_{\rm sym}\over6T}\,,\\
b&=6\left(2^{2/3}-1\right)+{-90S_0+30L-5K_{\rm sym}+Q_{\rm sym}\over2T}\,,\\
c&=-4\left(2^{2/3}-1\right)+{72S_0-24L+4K_{\rm sym}-Q_{\rm sym}\over2T}\,,\\
d&=2^{2/3}-1+{-60S_0+20L-3K_{\rm sym}+Q_{\rm sym}\over6T}\,.\end{split}
\label{eq:par}\end{align}
Because the value of $Q_0$ is quite uncertain we use $d^\prime=0$, which 
implies $Q_0=-432.3$ MeV for the typical
values $E_0=-16$ MeV, $K_0=220$ MeV, and $n_0=0.16$ fm$^{-3}$.  This
value matches the means of the values of $Q_n-Q_{\rm sym}$ for the
interactions displayed in Figures~\ref{fig:kn-l} and~\ref{fig:ksym-l}
for $L\sim50$ MeV.

%%%%%%%%%%%%%%%%%%%%%%%%%%%%%%%%
\begin{figure}[t]
\centering
\includegraphics[width=0.6\textwidth,angle=0]{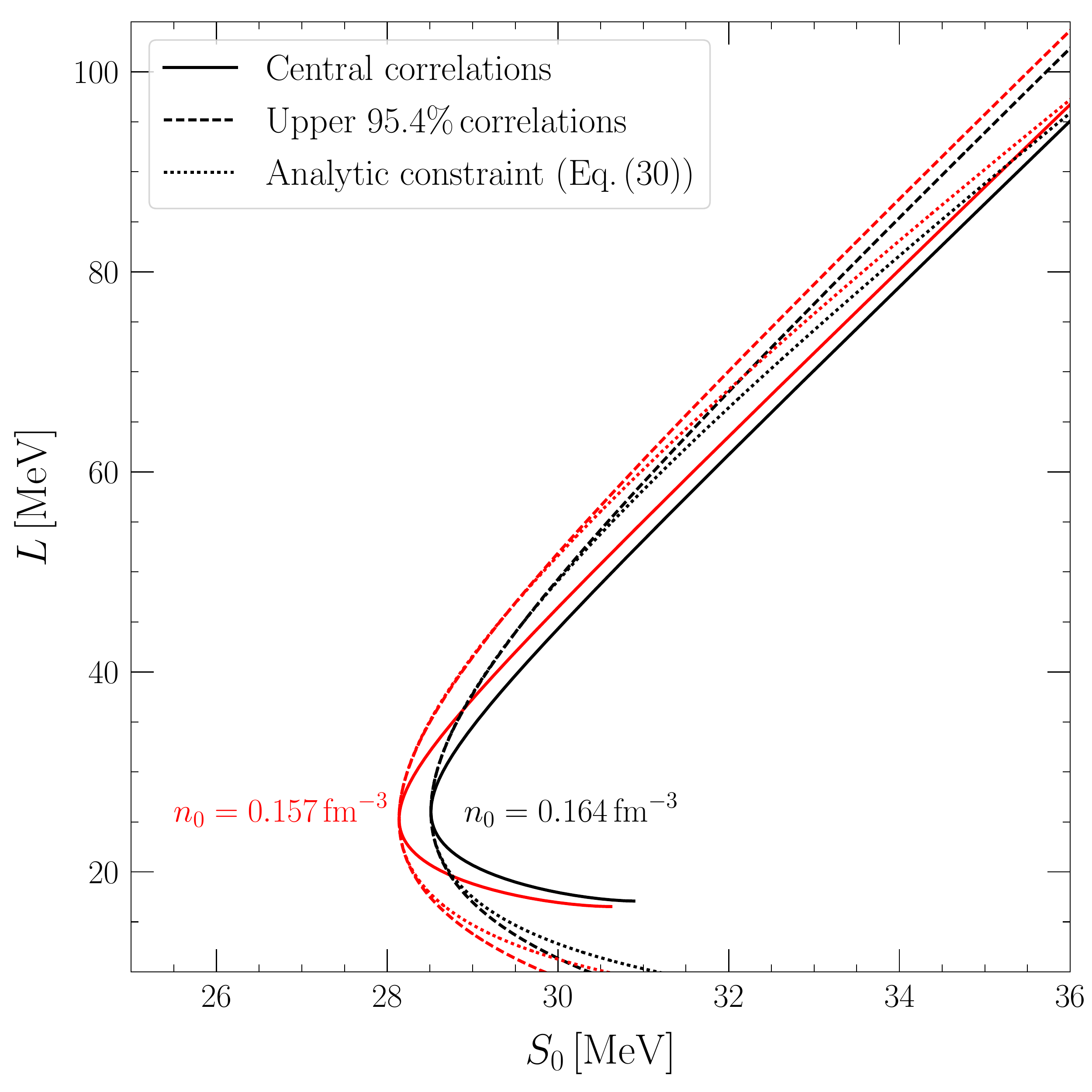}
\caption{Constraints on the symmetry parameters from the power-law
  energy-density functionals of Equations~\ref{eq:epar}) and~(\ref{eq:spar}) 
  are shown by solid and dashed black (red) curves for $n_0=0.164 (0.157)$ fm$^{-3}$ and $E_0=-15.5$ MeV.  The solid (dashed)
  curves correspond to the incorporation of the $K_n-L$ and $Q_n-L$
  central (upper 95\%) correlations. The black (red) dotted curves show the corresponding analytic
  constraints from Equation~(\ref{Eq.sl}).}\label{fig:powbound}
\end{figure}

%%%%%%%%%%%%%%%%%%%%%%%%%%%%%%%%

%%%%%%%%%%%%%%%%%%%%%%%%%%%%%%%%%%
\begin{figure}[t]%[thbp]
\centering
\includegraphics[width=0.6\textwidth,angle=0]{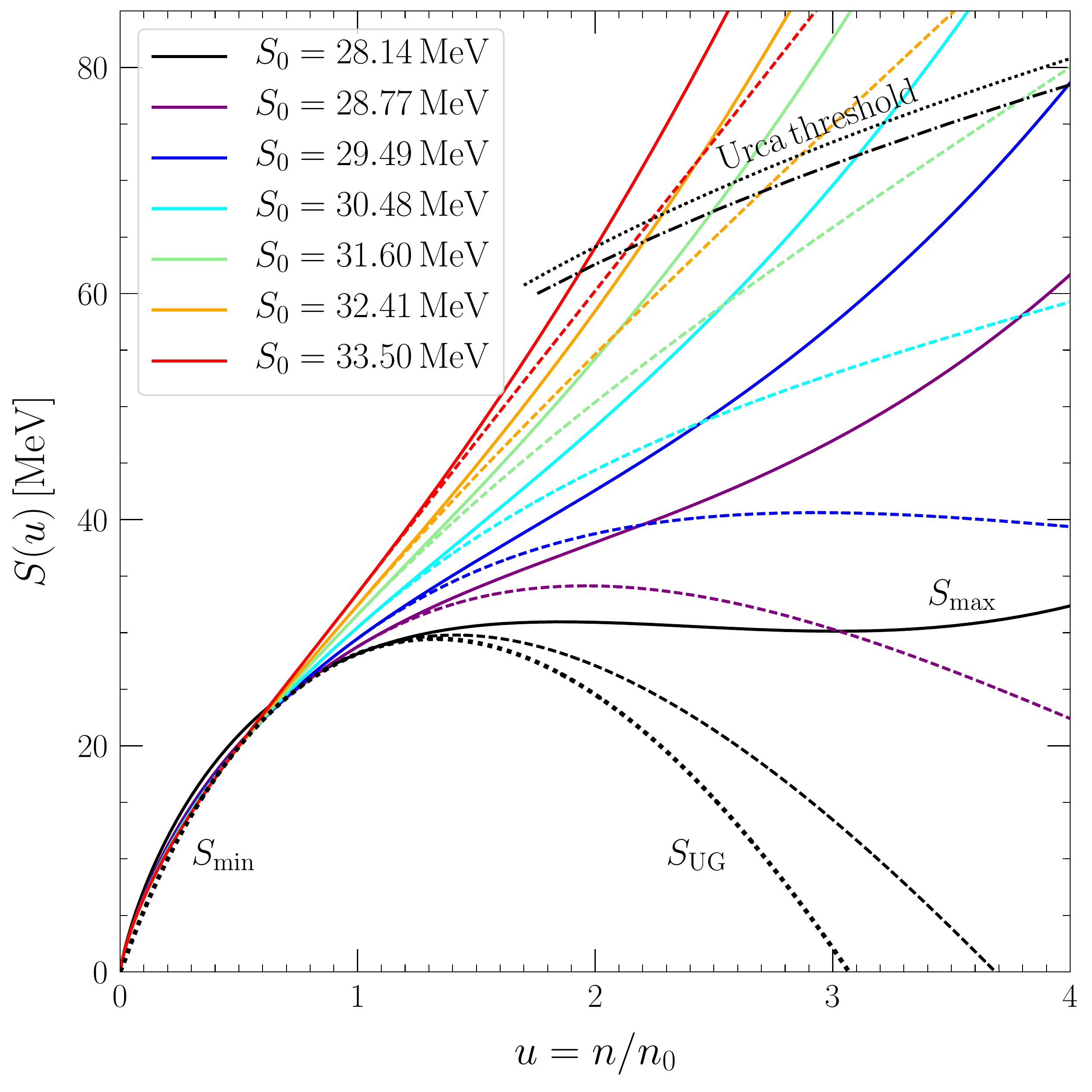}
\caption{Upper and lower limits to the symmetry energy from the
  UG constraint. Each solid (dashed) curve denotes the results
  for different choices of $S_0$ and uses the upper 95.4\% confidence 
  errors (central values) of the correlations between $K_{\rm sym}$ and $L$
  and between $Q_{\rm sym}$ and $L$.  For $u\ge1$, the solid curves
  denote $S_{\rm max}$, an effective upper limit to $S(u)$ as
  discussed in the text.  For $u\le1$, the solid curves denote an
  effective lower limit $S_{\rm min}$ and approach the UG
  bound (lower dotted curve) as $u\to0$.  The upper dotted (dotted-dashed) 
  curve indicates the threshold for the onset of the direct nucleon Urca
  process without (with) the inclusion of muons.}
\label{fig:smax}
\end{figure}

In this section, we always make use of this parametrization for the symmetry 
energy. While this parametrization allows the use of any reasonable 
value for the empirical parameters and thus does not automatically lead to 
any correlations between these parameters, this does not necessarily imply 
the absence of physical correlations between the parameters. These 
correlations, as suggested by other realistic parametrizations investigated in
Section~\ref{sec:rc},  are a result of correctly describing nuclear systems. 
In other words, some parameter choices in Equation~\eqref{eq:par} will lead to 
unphysical behavior. Thus in the following, we will utilize in addition the 
correlations between $K_{\rm sym}$, $Q_{\rm sym}$, and $L$. Then one 
can find the UG
bound $L(S_0)$ using the tangency conditions of Equation~(\ref{eq:tan}). For 
a given value of $u_t$, this results in two equations linear in $S_0$ and 
$L$. The most conservative bound is found utilizing the upper 95.4\% 
errors for these correlations -- see Figure~\ref{fig:powbound} -- and compares 
favorably with but is slightly less restrictive than the analytic bound from 
Equation~(\ref{Eq.sl}). Figure \ref{fig:powbound} also shows the dependence 
of the bounds on the assumed value of $n_0$.

We now discuss bounds on the symmetry energy as a function of density,
$S(u)$. The lower bound to $S(u)$ is just the UG bound and was
already presented in Figure~\ref{Fig:S_n}. Regarding the upper bound, we first 
consider the case when $u\ge1$. In this case, a conservative upper bound, 
$S_{\rm max}$, to $S(u)$ can be found for each value of $S_0$ by using the 
upper bound to $L$ for that $S_0$ and the upper 95.4\% confidence limits 
for the $K_{\rm sym}-L$ and $Q_{\rm sym}-L$ correlations from 
Equations~(\ref{eq:ksym-l}) and~(\ref{eq:qsym-l}). We show these $S_{\rm max}$ 
bounds for several values of $S_0$ in Figure~(\ref{fig:smax}). We also 
show a more realistic upper limit when the central values 
of the $K_{\rm sym}-L$ and $Q_{\rm sym}-L$ correlations are used.  
The same procedure leads to an effective lower limit to $S(u)$ for $u\le1$, 
$S_{\rm min}$, for each value of $S_0$. We stress that this 
effective lower limit is not a strict lower limit but a good approximation with 
desirable properties: as seen in Figure~\ref{fig:smax}, this 
procedure does not violate the UG bound, smoothly converges to 
the UG bound as $u\to0$, and has the correct noninteracting limiting 
behavior as $u\to0$. For $u\le1$ there is no practical upper bound to $S(u)$ 
because $L$ has a very small, and possibly negative, lower bound.

\subsection{Nuclear Surface Symmetry Energy}
Our UG bound has direct implications for the symmetry properties 
of the nuclear surface. For example, an effective maximum value of the 
surface symmetry energy parameter $S_S$ can be estimated by making 
the following argument.  We assume a potential model for the total free 
energy of semi-infinite symmetric matter in which the total energy density
$\mathcal{E}(u)$ is the sum of bulk and gradient contributions:
\begin{equation}
\mathcal{E}(u)=un_0E_{\rm SNM}(u)+Qn_0^2\left({du\over dz}\right)^2\,,
\end{equation}
where $z$ is the distance from the surface and $Q$ is a constant. Minimizing 
the total energy per unit surface area $\int_{-\infty}^\infty\mathcal{E}(u)dz$ with 
respect to $u(z)$ for a fixed number of baryons introduces the Lagrangian 
parameter $\mu$ and results in
\begin{equation}
u(E_{\rm SNM}(u)-\mu)=Qn_0\left({du\over dz}\right)^2\,,
\end{equation}
see \cite{Ravenhall83}.
One must have $\mu=E_{\rm SNM}(1)=E_0$ so that the gradient vanishes at the nuclear center ($u=1, z\to-\infty$).  A typical surface thickness parameter
\begin{equation}
t_{90-10}=\int_{0.1}^{0.9}{du\over(du/dz)}\equiv\sqrt{Qn_0\over T}I_t,\qquad I_t=\int_{0.1}^{0.9}{du\over\sqrt{f_B(u)}},
\label{eq:t9010}\end{equation}
can be defined where, phenomenologically, $t_{90-10}\simeq2.5$ fm and
\begin{equation}
f_B(u)=u{E_{\rm SNM}(u)-\mu\over T}=u\left[u^{2/3}+a^\prime u+ b^\prime u^{4/3}+c^\prime u^{5/3}+d^\prime u^2-{E_0\over T}\right].
\end{equation}

%%%%%%%%%%%%%%%%%%%%%%%%%%%%%%%%%%
\begin{figure}[t]%[thbp]
\centering
\includegraphics[width=0.6\textwidth,angle=0]{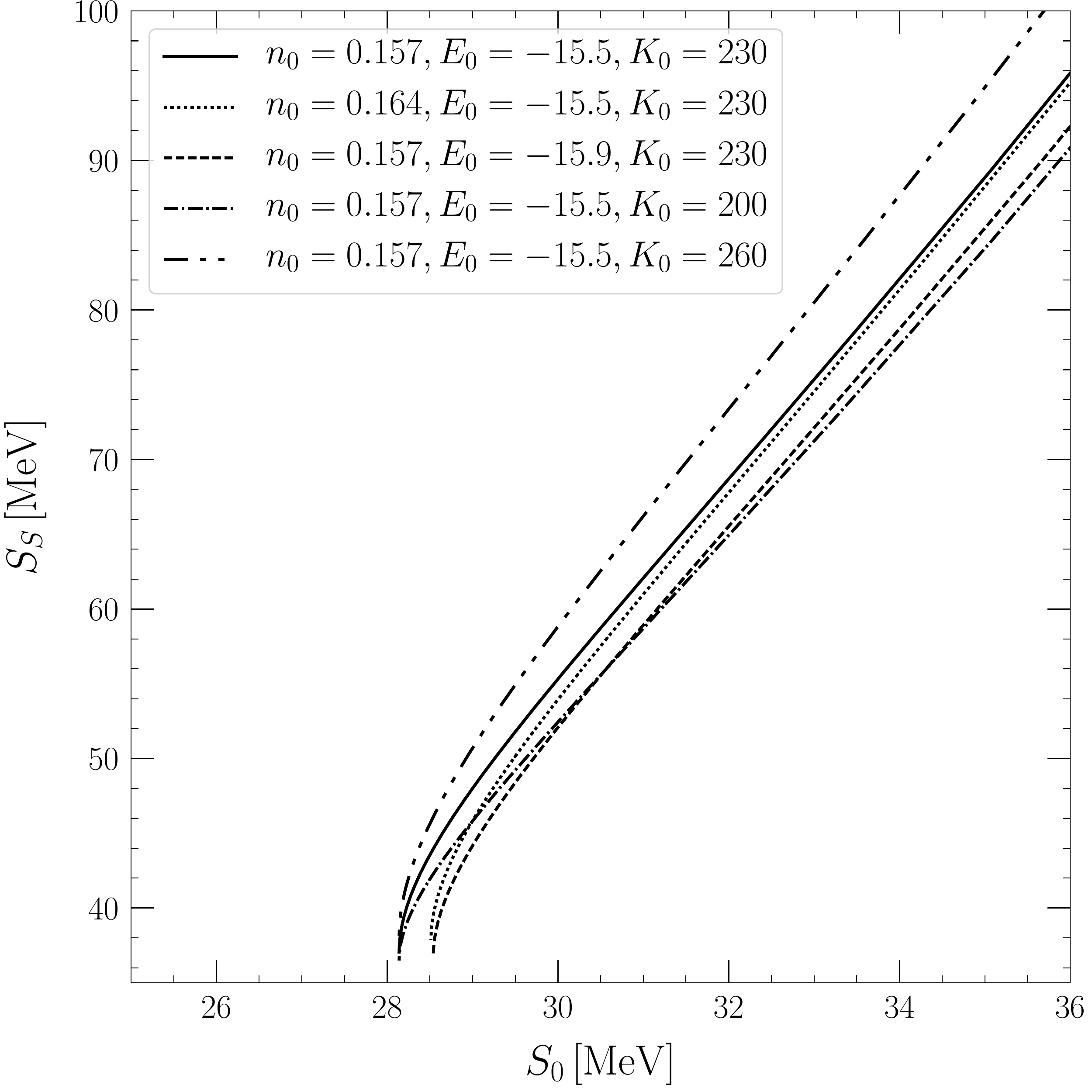}
\caption{The maximum surface symmetry parameter $S_S$ as a function 
of $S_0$.  The solid black curve assumes $n_0=0.157$ fm$^{-3}$, 
$E_0=-15.5$ MeV, and $K_0=230$ MeV.  Dotted (dashed) curves instead 
assume $n_0=0.164$ fm$^{-3}$ ($E_0=-15.9$ MeV).  The upper (lower) 
dashed-dotted curves assume $K_0=260 (200)$ MeV.}
\label{fig:surf}
\end{figure}

The surface tension of asymmetric matter can be expanded in terms of the 
neutron excess $\delta=1-2x$ at the center of the nucleus,
\begin{equation}
\sigma=\sigma_s-\sigma_\delta\delta^2,
\end{equation}
so the surface symmetry energy is $S_SA^{2/3}=4\pi r_o^2\sigma_\delta A^{2/3}$.  
It can be shown that
\begin{equation}
\sigma_\delta=S_0n_0\int_{-\infty}^\infty u\left({S_0\over S(u)}-1\right)dz\equiv S_0n_0\sqrt{Qn_0\over T}I_\delta,\qquad I_\delta=\int_0^1{u\over\sqrt{f_B(u)}}\left({S_0\over S(u)}-1\right)du,
\label{eq:ss}\end{equation}
which can also be written as $S_S=4\pi r_o^2n_0S_0t_{90-10}(I_\delta/I_t)$.  
Note that the integrand of
$I_\delta$ varies as $u^{-1/6}$ in the limit $u\to0$, which is why a
Taylor expansion of $S$ about $u=1$ should not be used to evaluate
this integral.  The surface symmetry energy scales roughly as
$K_0^{1/11}$, $|E_0|^{-1/8}$ and $n_0^{3/10}$.  It is clear that a maximum
value of the surface symmetry parameter can be determined if the
minimum allowed $S(u)$ for $u\le1$ is used in Equation~(\ref{eq:ss}) for
each value of $S_0$.  Figure~\ref{fig:surf} illustrates the resulting
maximum value of $S_S$, which for $S_0<33$ MeV has an upper limit of
about 80 MeV.

\subsection{The Crust-Core Boundary of Neutron Stars}

An approximate method of determining the location of the crust-core
boundary is to consider the stability of a homogeneous fluid of
nucleons in beta equilibrium.  \cite{BBP} showed that small density
fluctuations in an otherwise uniform fluid lead to instability when
\begin{equation}\label{df}
\mu_{pp}-\mu_{pn}^2\mu_{nn}^{-1}+4\sqrt{\pi \eta \alpha \hbar c}-4\eta\alpha(9\pi n_p^2)^{1/3}=0\,,
\end{equation}
where $\mu_{ij}=\partial\mu_i/\partial n_j$ and $\mu_i$ and $n_i$ are
the chemical potential and number density, respectively, of neutrons,
protons, or electrons. In the following, we use that 
$\mu_{pn}=\mu_{np}$ due to the commutativity of 
the derivative. The parameter $\alpha\simeq1/137$ is the 
fine-structure constant, and $\eta$ is determined by density-gradient terms 
in the nuclear Hamiltonian and can be approximated by
$\eta=D[1-4\mu_{np}/\mu_{nn}+(\mu_{np}/\mu_{nn})^2]$ where 
$D=4Q/3\simeq81$ MeV fm$^{5}$~\citep{HLPS13}. 
The chemical potentials and derivatives in Equation~(\ref{df}) are equivalent to
\begin{eqnarray}\label{df1}
\mu_n&=&{\partial nE\over\partial n}-x{\partial E\over\partial x},\qquad\mu_p={\partial nE\over\partial n}+(1-x){\partial E\over\partial x}\,,\cr
\mu_{nn}\mu_{pp}-\mu_{np}^2&=&{1\over n}{\partial^2nE\over\partial n^2}{\partial^2E\over\partial x^2}-\left({\partial^2E\over\partial n\partial x}\right)^2\,,\cr
\mu_{np}&=&{\partial^2nE\over\partial n^2}+(1-2x){\partial^2E\over\partial n\partial x}-{x(1-x)\over n}{\partial^2E\over\partial x^2}\,,\cr
\mu_{nn}&=&{\partial^2nE\over\partial n^2}-2x{\partial^2E\over\partial n\partial x}+{x^2\over n}{\partial^2E\over\partial x^2}\,.
\end{eqnarray}

%%%%%%%%%%%%%%%%%%%%%%%%%%%%%%%%%%
\begin{figure}[t]%[thbp]
\centering
\includegraphics[width=0.6\textwidth,angle=0]{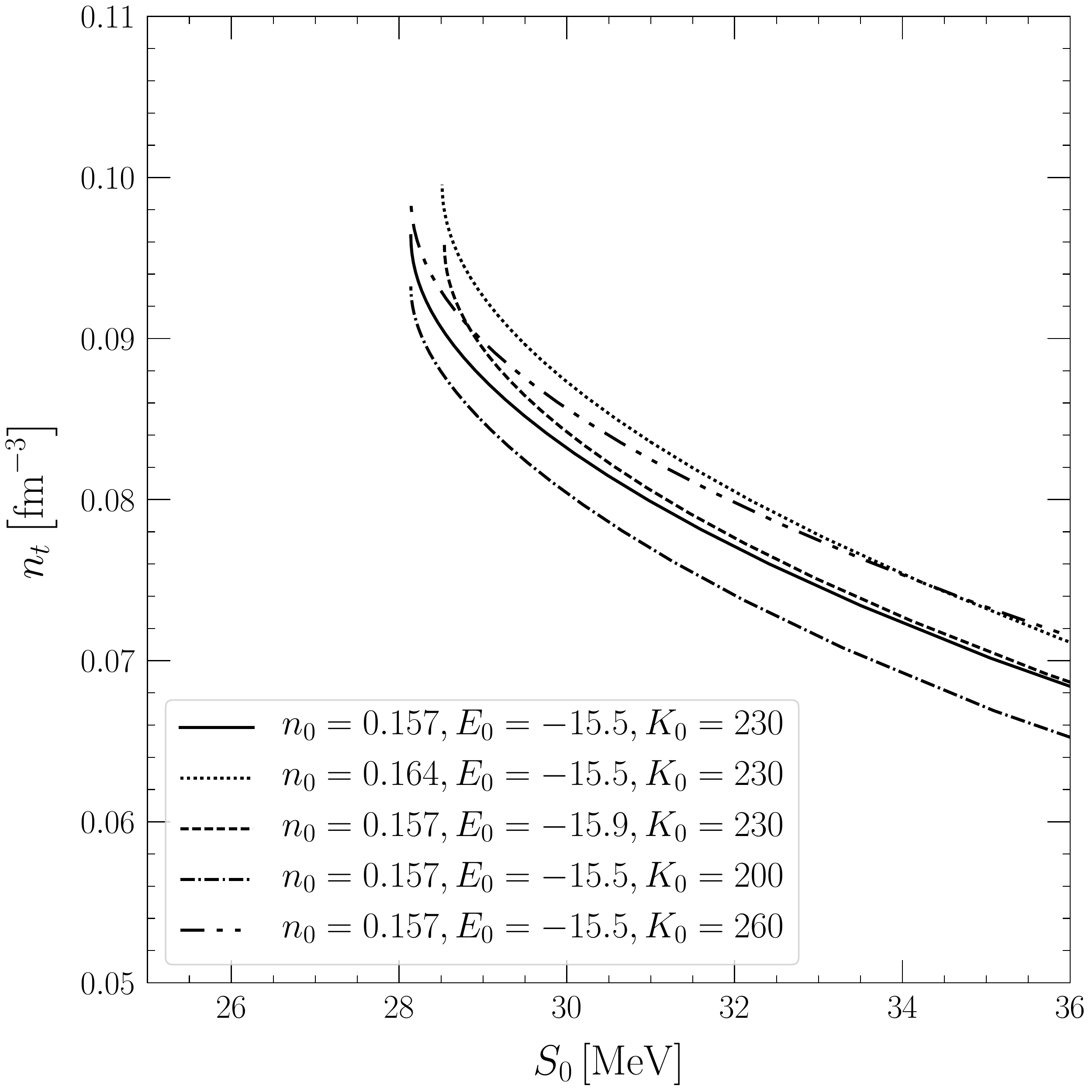}
\caption{The maximum value of the crust-core transition density as a function of $S_0$.  Curves are labeled as in Figure~\ref{fig:surf}.}
\label{fig:cc}
\end{figure}

We assume a quadratic symmetry energy so that $E(u,x)=E_{\rm SNM}(u)
+(1-2x)^2S(u)$.  The matter will be in $\beta$-equilibrium, for which 
$x(u)$ is determined by
 \begin{equation}
 \mu_n-\mu_p=4S(u)(1-2x)=\mu_e=\hbar c(3\pi^2n_0ux)^{1/3}\,.
 \label{eq:beta}\end{equation}  
At the densities of interest for the crust-core boundary, PNM 
is very similar to neutron-star matter since proton fractions are of the 
order 2.5-5\%.

If one employs the minimum value of $S(u)$ in the range $0<u<1$, the 
resulting crust-core transition density is maximized; see Figure~\ref{fig:cc}.  
We find that the crust-core transition density to decreases with an increasing 
$S_0$ and/or a decreasing $K_0$.  The transition density is roughly proportional
to the assumed value of $n_0$ but is insensitive to $E_0$ and is generally less
than 0.1 fm$^{-3}$.  This reinforces the notion that the transition density is 
larger than the approximate lower limit where nuclear pasta possibly exists,
$u\approx1/3$~\citep{Oyamatsu07}. 

\subsection{The Direct Urca Process}
The direct Urca process in neutron stars, in which the degenerate nucleons 
beta decay leading to subsequent cooling due to the loss of neutrinos and 
antineutrinos, requires a minimum fraction of protons, $x=1/9$, in the 
low-temperature limit and in the absence of muons due to energy and 
momentum conservation~\citep{Lattimer91}.  Assuming again the quadratic 
approximation for the dependence of the energy on the neutron excess, this 
condition, from Equation~(\ref{eq:beta}), is simply
\begin{equation}
{S(u)\over u^{1/3}}\ge{9\over28}\hbar c\left({\pi^2n_0\over3}\right)^{1/3}\simeq51.2~\mev\,,
\end{equation}
if the presence of muons is ignored. Muons appear around the nuclear
saturation density but do not change this condition significantly.

The threshold densities for the onset of the direct nucleon Urca process 
are shown in Figure~\ref{fig:smax}, for both the conservative and realistic 
maximum symmetry energies implied by the UG constraint. The 
Urca process is always disallowed for $u\simle2$ as long as 
$S_0<33.5$~MeV, irrespective of assumptions concerning the maximum 
symmetry energy.  For conservative (realistic) assumptions, and 
$S_0\simle29.5 (31.6)$ MeV, the Urca threshold density is $u\simge4$, 
near the maximum central densities of maximum-mass stars. This means 
that only the highest-mass neutron stars could become capable of rapid 
neutrino cooling due to the Urca process. Current estimates of the 
temperatures and ages of observed neutron stars are consistent with the 
interpretation that only a small fraction have cooled rapidly~\citep{Page04}.  
If mass is the most important parameter controlling neutron-star thermal 
evolution, then this small fraction implies that only very high-mass neutron 
stars cool rapidly,  providing empirical evidence that $S_0\simle32$ MeV.

\subsection{Neutron-star Radii and Neutron-skin Thicknesses}
The UG bounds directly impact predictions for both the neutron-star
radii and the neutron-skin thickness of neutron-rich nuclei, such
as $^{208}$Pb and $^{48}$ Ca.  From \cite{Lattimer13}, the correlation 
between $R_{1.4}$, the radius of a $1.4M_\odot$ neutron star, and the 
pressure of PNM at the saturation density ($p_{\rm PNM}(n_0)$) is
\begin{equation}
R_{1.4}\simeq(9.52\pm0.49)\left({fp_{\rm PNM}(n_0)\over\mev{\rm~fm}^{-3}}\right)^{1/4}{\rm~km}\,,
\end{equation}
where $f=0.94\pm0.02$ is a factor that corrects for the finite proton fraction 
of neutron-star matter.  Since $p_{\rm PNM}(n_0)=Ln_0/3$, one finds, using
$n_0=0.164\pm0.07$ fm$^{-3}$, that
\begin{equation}
R_{1.4}\simeq(4.51\pm0.26)\left({L\over\mev}\right)^{1/4}{\rm~km}\,.
\label{eq:r14}\end{equation}
With the upper limit $L\sim80$ MeV for $S_0\le33$ MeV, consistent with 
nuclear-mass measurements, we obtain $R_{1.4}<14.0$ km. The lower 
limit to $L$, $L_0$ for $S_0=S_0^{\rm LB}$, suggests that $R_{1.4}>9.5$ km.  
These limits are compatible with most estimates of neutron-star radii, but 
interpretations of currently proposed radius observations will probably not be 
significantly impacted by the UG constraint.

However, that is not the situation for the proposed neutron-skin thickness 
experiments.  \cite{Brown00} and \cite{Typel01} found that the neutron-skin 
thickness 
$r_{np}$ of $^{208}$Pb is related to $p_{\rm PNM}$ by
\begin{equation}
r_{np}\simeq(0.060\pm0.015)+0.12\left({p_{\rm PNM}(n=0.1{\rm~fm}^{-3})\over\mev{\rm~fm}^{-3}}\right){\rm~fm}\,.
\label{eq:rnp}\end{equation}
The minimum pressure of PNM, using the constraint 
$E_{\rm PNM}\ge E_{\rm UG}$, is
\begin{equation}
p_{\rm PNM, min}(n)={2\over3}E_{\rm UG}^0n\left({n\over n_0}\right)^{2/3},
%\begin{equation}
%p_{\rm PNM}(n)=u^2n_s\left[{L\over3}+{K_n\over9}(u-1)+{Q_n\over18}(u-1)^2\right].
\end{equation}
which is independent of assumptions about the value of $n_0$.   At the 
nominal density $n=0.1$ fm$^{-3}$, $p_{\rm PNM, min}=0.624~\mev{\rm~fm}^{-3}$.  

As previously mentioned, we cannot use either the Taylor expansions of
Equations~(\ref{Eq:ESNM}) and~(\ref{Eq:Sexp}) or the power-law expressions of
Equations~(\ref{eq:epar}) and~(\ref{eq:spar}) to estimate an upper limit to the 
energy or pressure of PNM because of the lack of an effective 
lower limit to $L$ as a function of $S_0$. However, the energy of PNM must be 
less than the Fermi-gas energy $T(2n/n_0)^{2/3}$ 
because the interactions are attractive at densities less than $n_0$.  This 
results in
\begin{equation}
p_{\rm PNM, max}(n)=n^2{\partial E_{\rm PNM, max}\over\partial n}={2\over3}T n \left({2n\over n_0}\right)^{2/3}=1.701~\mev{~\rm fm}^{-3},
%n^2{\partial(E_{\rm SNM}+S_{\rm max})\over\partial n}=Tn\left[{2\over3}\left({n\over n_0}\right)^{2/3}+(a+a^\prime){n\over n_}+{4\over3}(b+b^\prime)\left({n\over n_0}\right)^{4/3}+{5\over3}(c+c^\prime)\left({n\over n_0}\right)^{5/3}+2(d+d^\prime)\left({n\over n_0}\right)^2\right],
\end{equation}
which has been evaluated at the nominal density and is also independent 
of assumptions concerning $n_0$.  One therefore finds
0.12 fm$<r_{np}<0.28$ fm, independent of assumptions concerning $S_0$.  
While this is consistent with most experimental results for $^{208}$Pb, $r_{np}=0.159\pm0.041$ \citep{Danielewicz09}  (but see \cite{Danielewicz17}, 
who find $r_{np}=0.223\pm0.018$ fm), the most recent
neutron-skin thickness measurement from the PREX 
experiment~\citep{Abrahamyan12} is $r_{np}=0.33^{+0.16}_{-0.18}$ fm and 
thus the mean value is larger than our upper bound. The proposed 
PREX-II experiment will have 
an estimated error in $r_{np}$ of about $\pm0.06$ fm~\citep{Horowitz14}.  
This should be sufficient to resolve the tension with our upper limit.

\section{Conclusions}

We presented new bounds on the symmetry energy parameters $(S_0,L)$ 
based on the conjecture that the energy of the UG is less than 
the energy of PNM. Specifically, we determined a minimum
value for the volume symmetry parameter $S_0$ (equivalent to the
liquid drop parameter $J$), $S_0^{\rm LB}\simeq28.1$ MeV, as well as 
both minimum and maximum values of the symmetry energy slope 
parameter $L$ for values of $S_0\ge S_0^{\rm LB}$.  We also determined 
a minimum value for the symmetry incompressibility parameter 
$K_{\rm sym}$ as a function of $S_0$. These parameters are all 
evaluated at nuclear saturation density $n_0$, but in addition we  
established a minimum for the bulk symmetry energy $S(u)$ as a function 
of density in the vicinity of the saturation density $n_0$.

This conjecture is in agreement with ab initio calculations of PNM
with NN and 3N forces. Using conservative values for the Bertsch 
parameter $\xi_0$, as well as the saturation properties $E_0$, $n_0$, 
$K_0$, $K_{\rm sym}$, $Q_0$, and $Q_{\rm sym}$ of nuclear forces, 
we find that symmetry energy parameter 
constraints from various nuclear experiments are consistent with the 
UG bound.  However, several theoretical interactions in active use 
for both theoretical calculations of dense matter and tabulated equations 
of state used in astrophysical simulations of supernovae and neutron-star 
mergers violate the UG constraint.  In addition, less than 52\% of 
the more than 500 nonrelativistic potential and RMF interactions 
in recent compilations~\citep{Dutra12,Dutra14} satisfy our constraint.

Because the conjecture establishes a maximum value of $L$ for each $S_0$, 
and because both $K_{\rm sym}$ and the symmetry skewness parameter 
$Q_{\rm sym}$ are highly correlated with $L$, a maximum symmetry energy 
$S(n)$ for $n\simge n_0$ may be found for each assumed value of $S_0$.  
In addition, a minimum symmetry energy $S(n)$ for $n<n_0$ may also be 
obtained for each assumed value of $S_0$.  Our results thus have important 
consequences not only for astrophysical simulations but also for nuclear 
structure and neutron-star structure.  In particular, we obtain upper limits to 
the liquid-droplet surface symmetry energy parameter and the neutron-star 
crust-core transition density, and both upper and lower limits to the 
neutron-skin thickness of neutron-rich nuclei and typical neutron-star radii.  
We can also impose constraints on the operation of the nucleon direct Urca 
process in cooling neutron stars. 

Obtaining experimental results from cold atoms with a finite scattering
length~\citep{Hu10, Navon10, Horikoshi} will be important and may eventually 
offer additional insights and even more stringent constraints.

%%%%%%%%%%%%%%%%%%%%%%%%%%%%%%%%%%%%%%%%%%%%%%%%%%%%%%%%%%%%%%%%%%%%%%%%%%%%%%%%
\section*{Acknowledgements}
This work resulted from discussions at the YIPQS long-term and
Nishinomiya-Yukawa memorial workshop on Nuclear Physics, Compact
Stars, and Compact Star Mergers 2016 (YITP-T-16-02).
E.E.K., J.M.L., and I.T. thank the hospitality of the YITP.
The authors thank Stefano Gandolfi and other participants of the workshop
as well as Sanjay Reddy and Achim Schwenk for useful discussions,
Hajime Togashi for providing numerical data,
and Munekazu Horikoshi and Yoji Ohashi for inspiring suggestions.
This work was supported in part by
% IT
NSF Grant No. PHY-1430152 (JINA Center for the Evolution of the Elements),
% JML
US Department of Energy Grants DE-AC02-87ER40317 and DE-FG02-00ER41132,
% AO
%Grants-in-Aid from JSPS and MEXT (Nos. 15K05079, 15H03663, 16K05350, 24105001, 24105008),
JSPS/MEXT KAKENHI Grant Nos. 15K05079, 15H03663, 16K05350, 24105001, 24105008,
% EEK
and Slovak Grant VEGA-1/0469/15.

%%%%%%%%%%%%%%%%%%%%%%%%%%%%%%%%%%%%%%%%%%%%%%%%%%%%%%%%%%%%%%%%%%%%%%%%%%%%%%%%

%%%%%%%%%%%%%%%%%%%%%%%%%%%%%%%%%%%%%%%%%%%%%%%%%%%%%%%%%%%%%%%%%%%%%%%%%%%%%%%%
\end{document}